\documentclass[american,nofootinbib]{revtex4-1}
\usepackage[T1]{fontenc}
\usepackage{geometry}
\usepackage{color}
\usepackage{babel}
\usepackage{array}
\usepackage{verbatim}
\usepackage{wrapfig}
\usepackage{units}
\usepackage{multirow}
\usepackage{amsmath}
\usepackage{amssymb}
\usepackage{graphicx}
\usepackage[unicode=true,pdfusetitle,
 bookmarks=true,bookmarksnumbered=false,bookmarksopen=false,
 breaklinks=false,pdfborder={0 0 0},pdfborderstyle={},backref=false,colorlinks=true]
 {hyperref}
\hypersetup{
 linkcolor=coolblack,citecolor=burntorange,urlcolor=burntgreen}

\makeatletter

\providecommand{\tabularnewline}{\\}
\newcommand{\lyxdot}{.}

\usepackage{color}
\definecolor{burntorange}{rgb}{0.8, 0.33, 0.0}
\definecolor{charcoal}{rgb}{0.21, 0.27, 0.31}
\definecolor{coolblack}{rgb}{0.0, 0.28, 0.49}
\definecolor{burntgreen}{rgb}{0.05, 0.45, 0.27}

\makeatother

\begin{document}

\title{Efficiency of quantum versus classical annealing in non-convex learning
problems}

\author{Carlo Baldassi$^{1,2}$ and Riccardo Zecchina$^{1,3}$}

\affiliation{$^{1}$Bocconi Institute for Data Science and Analytics, Bocconi
University, Milano, Italy~\\
$^{2}$Istituto Nazionale di Fisica Nucleare, Sezione di Torino, Italy~\\
$^{3}$International Centre for Theoretical Physics, Trieste, Italy}
\begin{abstract}
Quantum annealers aim at solving non-convex optimization problems
by exploiting cooperative tunneling effects to escape local minima.
The underlying idea consists in designing a classical energy function
whose ground states are the sought optimal solutions of the original
optimization problem and add a controllable quantum transverse field
to generate tunneling processes. A key challenge is to identify classes
of non-convex optimization problems for which quantum annealing remains
efficient while thermal annealing fails. We show that this happens
for a wide class of problems which are central to machine learning.
Their energy landscapes is dominated by local minima that cause exponential
slow down of classical thermal annealers while simulated quantum annealing
converges efficiently to rare dense regions of optimal solutions.

\tableofcontents{}
\end{abstract}
\maketitle

\section{Introduction}

\begin{wrapfigure}{o}{0.5\columnwidth}%
\includegraphics[width=0.44\textwidth]{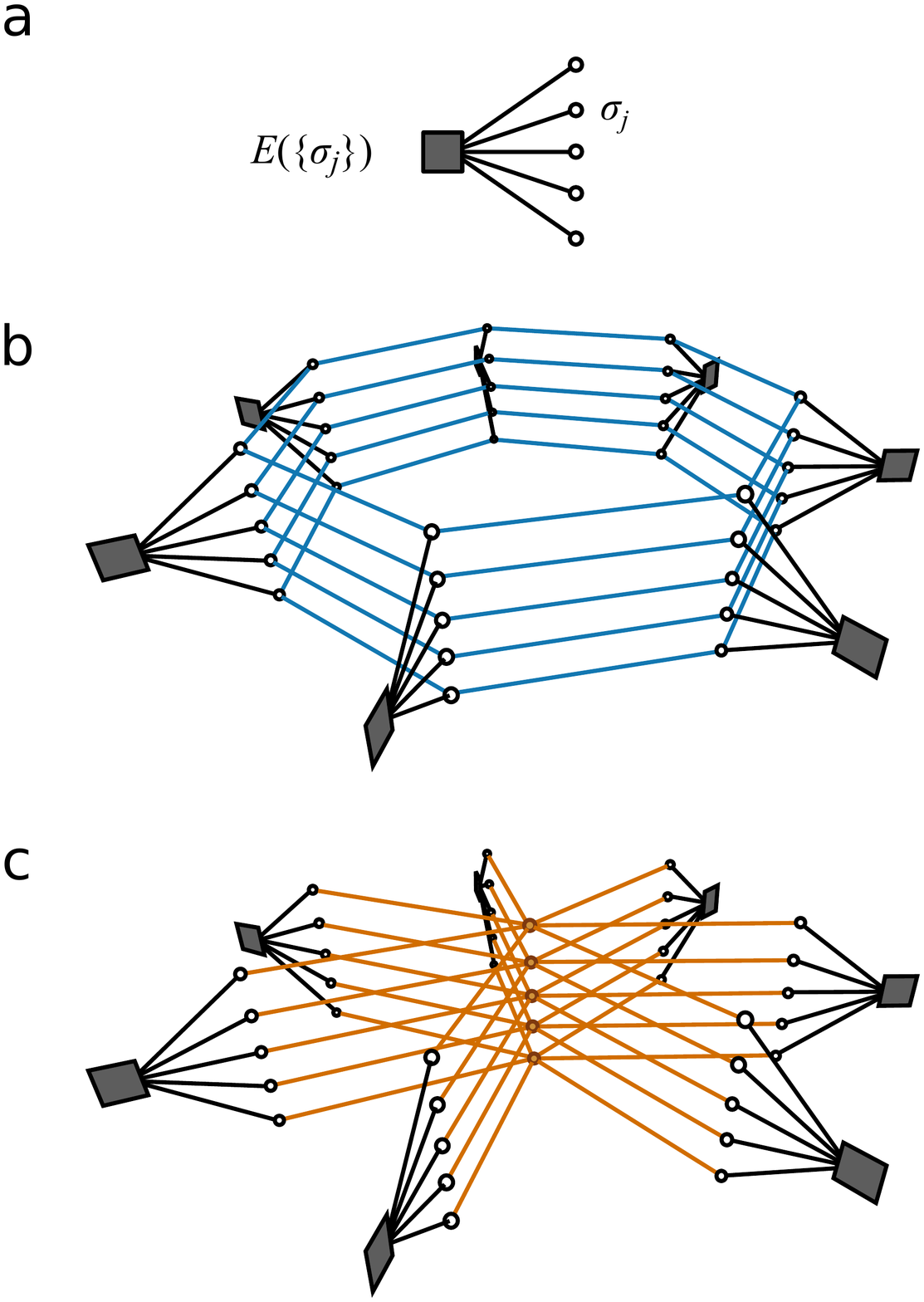}\caption{\label{fig:scheme}Topology of the Suzuki-Trotter vs Robust Ensemble
representations. \textbf{a:} the classical objective function we wish
to optimize which depends on $N$ discrete variables $\left\{ \sigma_{j}\right\} $
($N=5$ in the picture). \textbf{b:} Suzuki-Trotter interaction topology:
$y$ replicas of the classical system ($y=7$ in the picture) are
coupled by periodic 1 dimensional chains, one for each classical spin.
\textbf{c:} Robust Ensemble interaction topology: $y$ replicas are
coupled through a centroid configuration. In the limit of large $N$
and large $y$ (quantum limit) and for strong interaction couplings
all replicas are forced to be close, and the behavior of the two effective
models is expected to be similar.}
\end{wrapfigure}%

Quantum tunneling and quantum correlations govern the behavior of
very complex collective phenomena in quantum physics at low temperature.
Since the discovery of the factoring quantum algorithms in the 90s
\cite{shor1994algorithms}, a lot of efforts have been devoted to
the understanding of how quantum fluctuations could be exploited to
find low-energy configurations of energy functions which encode the
solutions of non-convex optimization problems in their ground states.
This has led to the notion of controlled quantum adiabatic evolution,
where a  time dependent many-body quantum system is evolved towards
its ground states so as to escape local minima through multiple tunneling
events \cite{ray1989sherrington,finnila1994quantum,kadowaki1998quantum,farhi2001quantum,das2008colloquium}.
When finite temperature effects have to be taken into account, the
computational process is called Quantum Annealing (QA). Classical
Simulated Annealing (SA) uses thermal fluctuations for the same computational
purpose, and Markov Chains based on this principle are among the most
widespread optimization techniques across science \cite{moore2011nature}.
Quantum fluctuations are qualitatively different from thermal fluctuations
and in principle quantum annealing algorithms could lead to extremely
powerful alternative computational devices.

In the quantum annealing approach, a time dependent quantum transverse
field is added to the classical energy function leading to an interpolating
Hamiltonian that may take advantage of correlated fluctuations mediated
by tunneling. Starting with a high transverse field, the quantum model
system can be initialized in its ground state, e.g. all spins aligned
in the direction of the field. The adiabatic theorem then ensures
that by slowly reducing the transverse field the system remains in
the ground state of the interpolating Hamiltonian. At the end of the
process the transverse field vanishes and the systems ends up in the
sought ground state of the classical energy function. The original
optimization problem would then be solved if the overall process could
take place in a time bounded by some low degree polynomial in the
size of the problem. Unfortunately, the adiabatic process can become
extremely slow. The adiabatic theorem requires the rate of change
of the Hamiltonian to be smaller than the square of the gap between
the ground state and the first excited state \cite{born1928beweis,landau1932theorie,zener1932non}.
For small gaps the process can thus become inefficient. Exponentially
small gaps are not only possible in worst case scenarios but have
also been found to exist in typical random systems where comparative
studies between quantum and classical annealing have so far failed
in displaying quantum exponential speed up, e.g. at first order phase
transition in quantum spin glasses \cite{altshuler2010anderson,bapst2013quantum}
or 2D spin glass systems \cite{santoro2002theory,martovnak2002quantum,heim2015quantum}.
More positive results have been found for ad hoc energy functions
in which global minima are planted in such a way that tunneling cascades
can become more efficient than thermal fluctuations \cite{ronnow2014defining,farhi2001quantum}.
As far as the physical implementations of quantum annealers is concerned,
studies have been focused on discriminating the presence of quantum
effects rather than on their computational effectiveness \cite{johnson2011quantum,boixo2014evidence,langbein2004control}.

Consequently, a key open question is to identify classes of relevant
optimization problems for which quantum annealing can be shown to
be exponentially faster than its classical thermal counterpart.

Here we give an answer to this question by providing analytic and
simulation evidence of exponential speed up of quantum versus classical
simulated annealing for a representative class of random non-convex
optimization problems of basic interest in machine learning. The simplest
example of this class is the problem of training binary neural networks
(described in detail below): very schematically, the variables of
the problem are the (binary) connection weights, while the energy
measures the training error over a given dataset.

These problems have been very recently found to possess a rather distinctive
geometrical structure of ground states \cite{baldassi_subdominant_2015,baldassi_unreasonable_2016,baldassi_local_2016,baldassi2016learning}:
the free energy landscape has been shown to be characterized by the
existence of an exponentially large number metastable states and isolated
ground states, and a few regions where the ground states are dense.
These dense regions, which had previously escaped the equilibrium
statistical physics analysis \cite{krauth-mezard,sompolinsky1990learning},
are exponentially rare, but still possess a very high local internal
entropy: they are composed of ground states that are surrounded, at
extensive but relatively small distances, by exponentially many other
ground states. Under these circumstances, classical SA (as any Markov
Chain satisfying detailed balance) gets trapped in the metastable
states, suffering ergodicity breaking and exponential slowing down
toward the low energy configurations. These problems have been considered
to be intractable for decades and display deep similarities with disordered
spin glass models which are known to never reach equilibrium.

The large deviation analysis that has unveiled the existence of the
rare dense regions has led to several novel algorithms, including
a Monte Carlo scheme defined over an appropriate objective function
\cite{baldassi_unreasonable_2016} that bears close similarities with
a Quantum Monte Carlo (QMC) technique based on the Suzuki-Trotter
transformation \cite{das2008colloquium}. Motivated by this analytical
mapping and by the geometrical structure of the dense and degenerate
ground states which is expected to favor zero temperature kinetic
processes \cite{foini2010solvable,biroli2012tentative}, we have conducted
a full analytical and numerical statistical physics study of the quantum
annealing problem, reaching the conclusion that in the quantum limit
the QMC process, i.e.~Simulated Quantum Annealing (SQA), can equilibrate
efficiently while the classical SA gets stuck in high energy metastable
states. These results generalize to multi layered networks.

While it is known that other quasi-optimal classical algorithms for
the same problems exist \cite{baldassi_unreasonable_2016,hubara2016quantized,courbariaux2015binaryconnect},
here we focus on the physical speed up that a quantum annealing approach
could  provide in finding rare regions of ground states. We provide
physical arguments and numerical results supporting the conjecture
that the real time quantum annealing dynamics behaves similarly to
SQA.

As far as machine learning is concerned, dense regions of low energy
configurations (i.e.~quasi-flat minima over macroscopic length scales)
are of fundamental interest, as they are particularly well-suited
for making predictions given the learned data: on the one hand, these
regions are by definition robust with respect to fluctuations in a
sizable fraction of the weight configurations and as such are less
prone to fit the noise. On the other hand, an optimal Bayesian estimate,
resulting from a weighted consensus vote on all configurations, would
receive a major contribution from one of such regions, compared to
a narrow minimum; the centroid of the region (computed according to
any reasonable metric which correlates the distance between configurations
with the network outcomes) would act as a representative of the region
as a whole \cite{mackay2003information}. In this respect, it is worth
mentioning that in deep learning \cite{lecun2015deep} all the learning
algorithms which lead to good prediction performance always include
effects of a systematically injected noise in the learning phase,
a fact that makes the equilibrium Gibbs measure not the stationary
measure of the learning protocols and drive the systems towards wide
minima. We expect that these results can be generalized to many other
classes of non convex optimization problems where local entropy plays
a role, ranging from robust optimization to physical disordered systems.

Quantum gate based algorithms for machine learning exist, however
the possibility of a physical implementation remains a critical issue
\cite{aaronson2015read}.

\section{Energy functions}

As a working example, we first consider the problem of learning random
patterns in single layer neural network with binary weights, the so
called binary perceptron problem \cite{krauth-mezard}. This network
maps vectors of $N$ inputs $\xi\in\left\{ -1,+1\right\} ^{N}$ to
binary outputs $\tau=\pm1$ through the non linear function $\tau=\mathrm{sgn}\left(\sigma\cdot\xi\right)$,
where $\sigma\in\left\{ -1,+1\right\} ^{N}$ is the vector of synaptic
weights. Given $\alpha N$ input patterns $\left\{ \xi^{\mu}\right\} _{\mu=1}^{\alpha N}$
with $\mu=1,...,\alpha N$ and their corresponding desired outputs
$\left\{ \tau^{\mu}\right\} _{\mu=1}^{\alpha N}$, the learning problem
consists in finding $\sigma$ such that all input patterns are simultaneously
classified correctly, i.e.~$\mathrm{sgn}\left(\sigma\cdot\xi^{\mu}\right)=\tau^{\mu}$
for all $\mu$. Both the components of the input vectors $\xi_{i}^{\mu}$
and the outputs $\tau^{\mu}$ are independent identically distributed
unbiased random variables ($P\left(x\right)=\frac{1}{2}\delta\left(x-1\right)+\frac{1}{2}\delta\left(x+1\right)$).
In the binary framework, the procedure for writing a spin Hamiltonian
whose ground states are the sought optimal solutions of the original
optimization problem is well known \cite{barahona1982computational}.
The energy $E$ of the binary perceptron is proportional to the number
of classification errors and can be written as
\begin{equation}
E\left(\left\{ \sigma_{j}\right\} \right)=\sum_{\mu=1}^{\alpha N}\Delta_{\mu}^{n}\Theta\left(-\Delta_{\mu}\right),\ \ \ \ \Delta_{\mu}\doteq\frac{\tau^{\mu}}{\sqrt{N}}\sum_{j=1}^{N}\xi_{j}^{\mu}\sigma_{j}\label{eq:E}
\end{equation}
where $\Theta\left(x\right)$ is the Heaviside step function: $\Theta\left(x\right)=1$
if $x>0$, $\Theta\left(x\right)=0$ otherwise. When the argument
of the $\Theta$ function is positive, the perceptron is implementing
the wrong input-output mapping. The exponent $n\in\left\{ 0,1\right\} $
defines two different forms of the energy functions which have the
same zero energy ground states and different structures of local minima.
The equilibrium analysis of the binary perceptron problem shows that
in the large size limit and for $\alpha<\alpha_{c}\simeq0.83$ \cite{krauth-mezard},
the energy landscape is dominated by an exponential number of local
minima and of zero energy ground states that are typically geometrically
isolated \cite{huang2014origin}, i.e.~they have extensive mutual
Hamming distances. For both choices of $n$ the problem is computationally
hard for SA processes \cite{horner1992dynamics}: in the large $N$
limit, a detailed balanced stochastic search process gets stuck in
metastable states at energy levels of order $O(N)$ above the ground
states.

Following the standard SQA approach, we identify the binary variables
$\sigma$ with one of the components of physical quantum spins, say
$\sigma^{z}$, and we introduce the Hamiltonian operator of a model
of $N$ quantum spins with the perceptron term of Eq.~(\ref{eq:E})
acting in the longitudinal direction $z$ and a magnetic field $\Gamma$
acting in the transverse direction $x$. The interpolating Hamiltonian
reads:
\begin{equation}
\hat{H}=E\left(\left\{ \hat{\sigma}_{j}^{z}\right\} \right)-\Gamma\sum_{j=1}^{N}\hat{\sigma}_{j}^{x}\label{eq:H}
\end{equation}
where $\hat{\sigma}_{j}^{z}$ and $\hat{\sigma_{j}}^{x}$ are the
spin operators (Pauli matrices) in the $z$ and $x$ directions. For
$\Gamma=0$ one recovers the classical optimization problem. The QA
procedure consists in initializing the system at large $\beta$ and
$\Gamma$, and slowly decreasing $\Gamma$ to $0$. To analyze the
low temperature phase diagram of the model we need to study the average
of the logarithm of the partition function $Z=\mathrm{Tr}\,\left(e^{-\beta\hat{H}}\right).$
This can be done using the Suzuki-Trotter transformation which leads
to the study of a classical effective Hamiltonian acting on a system
of $y$ interacting Trotter replicas of the original classical system
coupled in an extra dimension:
\begin{equation}
H_{\mathrm{eff}}\left(\left\{ \sigma_{j}^{a}\right\} _{j,a}\right)=\frac{1}{y}\sum_{a=1}^{y}E\left(\left\{ \sigma_{j}^{a}\right\} _{j}\right)-\frac{\gamma}{\beta}\sum_{a=1}^{y}\sum_{j=1}^{N}\sigma_{j}^{a}\sigma_{j}^{a+1}-\frac{NK}{\beta}\label{eq:E-eff}
\end{equation}
where the $\sigma_{j}^{a}=\pm1$ are Ising spins, $a\in\left\{ 1,\dots,y\right\} $
is a replica index with periodic boundary conditions $\sigma_{j}^{y+1}\equiv\sigma_{j}^{1}$,
$\gamma=\frac{1}{2}\log\coth\left(\frac{\beta\Gamma}{y}\right)$ and
$K=\frac{1}{2}y\log\left(\frac{1}{2}\sinh\left(2\frac{\beta\Gamma}{y}\right)\right).$

The replicated system needs to be studied in the limit $y\to\infty$
to recover the so called path integral continuous quantum limit and
to make the connection with the behavior of quantum devices \cite{heim2015quantum}.
The SQA dynamical process samples configurations from an equilibrium
distribution and it is not necessarily equivalent to the real time
Schr\"{o}dinger equation evolution of the system. A particularly dangerous
situation occurs if the ground states of the system encounter first
order phase transitions which are associated to exponentially small
gaps \cite{altshuler2010anderson,bapst2012quantum,bapst2013thermal}
at finite N. As discussed below, this appears not to be the case for
the class of models we are considering.

\section{Connection with the local entropy measure}

The effective Hamiltonian Eq.~(\ref{eq:E-eff}) can be interpreted
as many replicas of the original systems coupled through one dimensional
periodic chains, one for each original spin, see Fig.~\ref{fig:scheme}b.
Note that the interaction term $\gamma$ diverges as the transverse
field $\Gamma$ goes to $0$. This geometrical structure is very similar
to that of the Robust Ensemble (RE) formalism \cite{baldassi_unreasonable_2016},
where a probability measure that gives higher weight to rare dense
regions of low energy states is introduced. There, the main idea is
to maximize $\Phi\left(\sigma^{\star}\right)=\log\sum_{\left\{ \sigma\right\} }e^{-\beta E\left(\sigma\right)-\lambda\sum_{j=1}^{N}\sigma_{j}\sigma_{j}^{\star}}$,
i.e. a ``local free entropy'' where $\lambda$ is a Lagrange parameter
that controls the extensive size of the region around a reference
configuration $\sigma^{\star}$. One can then build a new Gibbs distribution
$P\left(\sigma^{\star}\right)\propto e^{y\Phi\left(\sigma^{\star}\right)}$,
where $-\Phi$ has the role of an energy and $y$ of an inverse temperature:
in the limit of large $y$, this distribution concentrates on the
maxima of $\Phi$. Upon restricting the values of $y$ to be integer
(and large), $P\left(\sigma^{\star}\right)$ takes a factorized form
yielding a replicated probability measure $P_{\mathrm{RE}}\left(\sigma^{\star},\sigma^{1},\dots,\sigma^{y}\right)\propto e^{-\beta H_{\mathrm{eff}}^{\mathrm{RE}}\left(\sigma^{\star},\left\{ \sigma_{j}^{a}\right\} \right)}$
where the effective energy is given by 
\begin{equation}
H_{\mathrm{eff}}^{\mathrm{RE}}\left(\sigma^{\star},\left\{ \sigma_{j}^{a}\right\} _{j,a}\right)=\sum_{a=1}^{y}E\left(\left\{ \sigma_{j}^{a}\right\} _{j}\right)-\frac{\lambda}{\beta}\sum_{a=1}^{y}\sum_{j=1}^{N}\sigma_{j}^{a}\sigma_{j}^{\star}\label{eq:H_RE}
\end{equation}

As in the Suzuki-Trotter formalism, $H_{\mathrm{eff}}^{\mathrm{RE}}\left(\sigma^{\star},\left\{ \sigma_{j}^{a}\right\} _{j,a}\right)$
corresponds to a system with an overall energy given by the sum of
$y$ individual ``real replica energies'' plus a geometric coupling
term; in this case however the replicas interact with the ``reference''
configurations $\sigma^{\star}$ rather than among themselves, see
Fig.~\ref{fig:scheme}c.

The Suzuki-Trotter representation and the RE formalism differ in the
topology of the interactions between replicas and in the scaling of
the interactions, but for both cases there is a classical limit, $\Gamma\to0$
and $\lambda\to\infty$ respectively, in which the replicated systems
are forced to correlate and eventually coalesce in identical configurations.
For non convex problems, these will not in general correspond to configuration
dominating the original classical Gibbs measure.

For the sake of clarity we should remind that in the classical limit
and for $\alpha<\alpha_{c}$, our model presents an exponential number
of far apart isolated ground states which dominate the Gibbs measure.
At the same time, there exist rare clusters of ground states with
a density close to its maximum possible value (high local entropy)
for small but still macroscopic cluster sizes \cite{baldassi_subdominant_2015}.
This fact has several consequences: no further subdivision of the
clusters into states is possible, the ground states are typically
$O(1)$ spin flip connected \cite{baldassi_subdominant_2015} and
a tradeoff between tunneling events and exponential number of destination
states within the cluster is possible.

\section{Phase diagram: analytical and numerical results}

Thanks to the mean field nature of the energetic part of the system,
Eq.~(\ref{eq:E-eff}), we can resort to the replica method for calculating
analytically the phase diagram. As discussed in the Appendix Sec.~\ref{sec:analysis},
this can be done under the so called static approximation, which consists
in using a single parameter $q_{1}$ to represent the overlaps along
the Trotter dimension, $q_{1}^{ab}=\left\langle \frac{1}{N}\sum_{j=1}^{N}\sigma_{j}^{a}\sigma_{j}^{b}\right\rangle \approx q_{1}$.
Although this approximation crudely neglects the dependency of $q_{1}^{ab}$
from $\left|a-b\right|$, the resulting predictions show a remarkable
agreement with numerical simulations.

\begin{figure}
\begin{centering}
\includegraphics[width=1\textwidth]{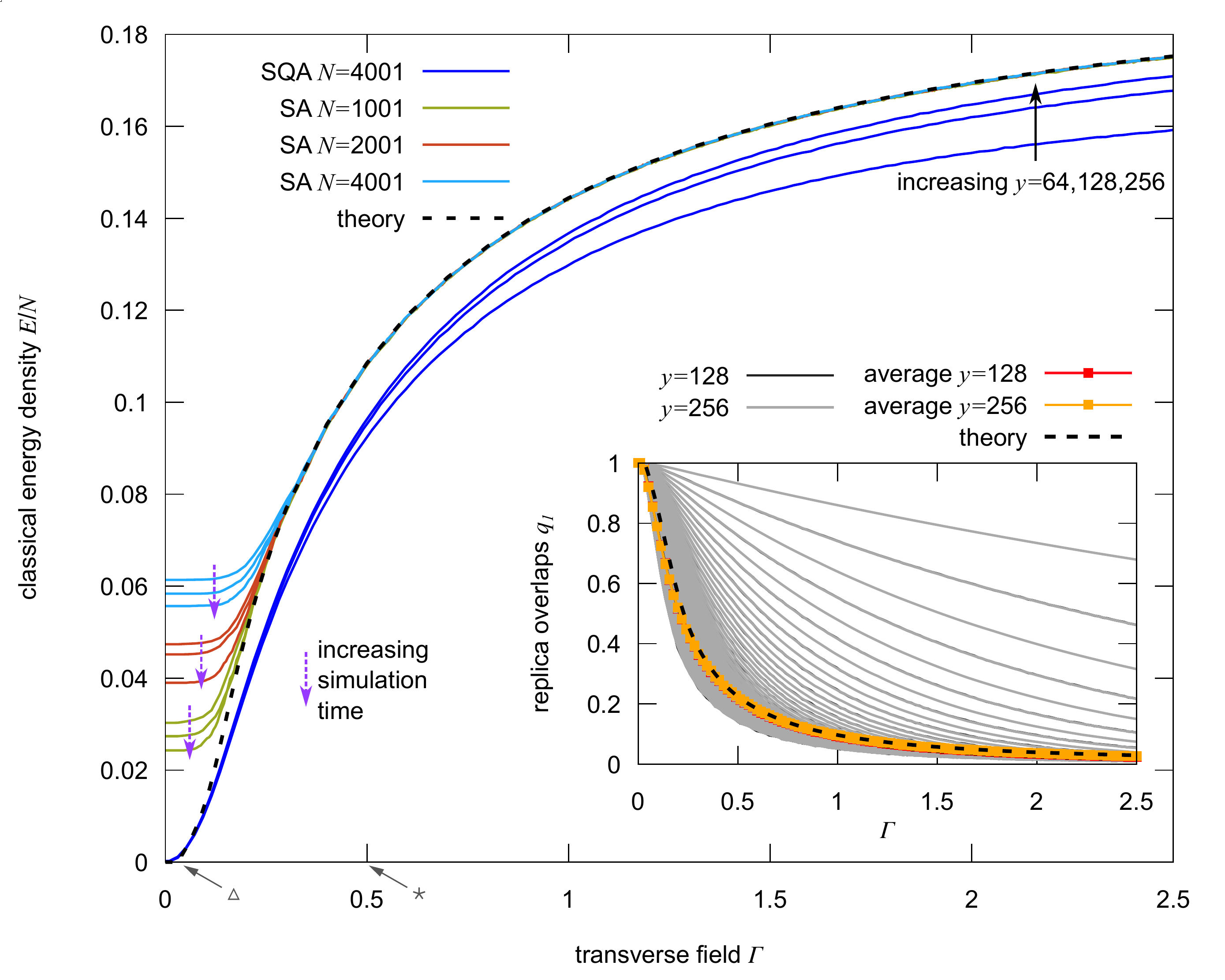}
\par\end{centering}
\caption{\label{fig:sim_vs_theory}Classical energy density (i.e.~longitudinal
component of the energy, divided by $N$) as a function of the transverse
field $\Gamma$ (single layer problems with $\alpha=0.4$ and $n=0$,
$15$ independent samples per curve). The QA simulations at $\beta=20$
approach the theoretical prediction as $y$ increases (cf.~black
arrow). The results do not change significantly when varying $N$
or the simulation time (the curves with $N=1001$ or $N=2001$ are
indistinguishable from the ones displayed at this level of detail).
All SA simulations instead got stuck and failed to equilibrate at
low enough temperatures (small equivalent $\Gamma$). The results
are noticeably worse for larger $N$, and doubling or quadrupling
the simulation time doesn't help much (cf.~purple arrows). \emph{Inset:}
Trotter replicas overlaps $q_{1}^{ab}$ (same data as for the main
figure). The theoretical prediction is in remarkably good agreement
with the average value measured from the simulations (the $y=128$
curve is barely visible under the $y=256$ one). The gray curves show
the overlaps at varying distances along the Trotter dimension: the
topmost one is the overlap between neighboring replicas $q_{1}^{a\left(a+1\right)}$,
then there is the overlap between second-neighbors $q_{1}^{a\left(a+2\right)}$
and so on (cf.~Fig.~\ref{fig:scheme}). The $y=128$ curves are
essentially hidden under the $y=256$ ones and can only be seen from
their darker shade, following an alternating pattern.}
\end{figure}

In the main panel of Fig.~\ref{fig:sim_vs_theory}, we report the
analytical predictions for the average classical component of the
energy of the quantum model as a function of the transverse field
$\Gamma$. We compare the results with the outcome of extensive simulations
performed with the reduced-rejection-rate Monte Carlo method \cite{baldassi_method_2017},
in which $\Gamma$ is initialized at $2.5$ and gradually brought
down to $0$ in regular small steps, at constant temperature, and
fixing the total simulation time to $\tau Ny\cdot10^{4}$ (as to keep
constant the number of Monte Carlo sweeps when varying $N$ and $y$).
The details are reported in the Appendix Sec.~\ref{sec:Numerical-simulations-details}.
The size of the systems, the number of samples and the number of Trotter
replicas are scaled up to large values so that both finite size effects
and the quantum limit are kept under control. A key point  is to observe
that the results do not degrade with the number of Trotter replicas:
the average ground state energy approaches a limiting value, close
to the theoretical prediction, in the large $y$ quantum limit. The
results appear to be rather insensitive to both $N$ and the simulation
time scaling parameter $\tau$. This indicates that Monte Carlo appears
to be able to equilibrate efficiently, in a constant (or almost constant)
number of sweeps, at each $\Gamma$. The analytical prediction for
the classical energy only appears to display a relatively small systematic
offset (due to the static approximation) at intermediate values of
$\Gamma$, while it is very precise at both large and small $\Gamma$;
the expectation of the total Hamiltonian on the other hand is in excellent
agreement with the simulations (see Appendix Sec.~\ref{sec:Numerical-simulations-details}).

In the same plot we display the behavior of classical SA simulated
with a standard Metropolis-Hastings scheme, under an annealing protocol
in $\beta$ that would follow the same theoretical curve as SQA if
the system were able to equilibrate (see Appendix Sec.~\ref{sec:Numerical-simulations-details}):
as expected \cite{horner1992dynamics}, SA gets trapped at very high
energies (increasing with problem size; in the thermodynamic limit
it is expected that SA would remain stuck at the initial value $0.5N$
of the energy for times which scale exponentially with $N$). Alternative
annealing protocols yield analogous results; the exponential scaling
with $N$ of SA on binary perceptron models had also been observed
experimentally in previous results, e.g.~in refs.~\cite{baldassi_local_2016,baldassi-et-all-pnas}.

In the inset of Fig.~\ref{fig:sim_vs_theory} we report the analytical
prediction for the transverse overlap parameter $q_{1}$, which quite
remarkably reproduces fairly well the average overlap as measured
from simulations.

\begin{figure*}
\includegraphics[width=1\textwidth]{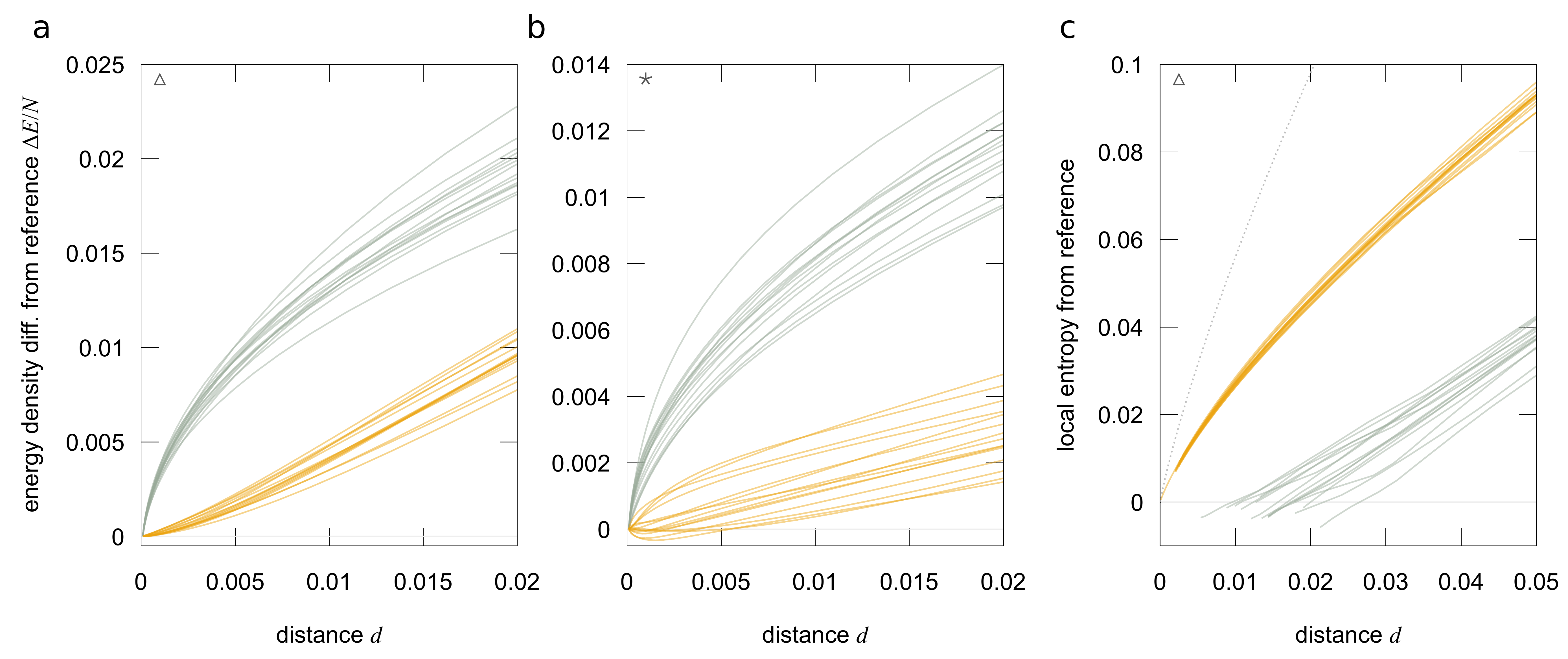}

\caption{\label{fig:en_landscapes}Panels \textbf{a} and \textbf{b}: energetic
profiles (in terms of the classical energy $E$, Eq.~(\ref{eq:E}))
around the configurations reached during the annealing process, comparing
QA (orange lower curves) with SA (gray top curves). The profiles represent
the most probable value of the energy density shift $\Delta E/N$
with respect to the reference point when moving away from the reference
at a given normalized Hamming distance $d$. The curves refer to the
data shown in Fig.~\ref{fig:sim_vs_theory}, using two different
times in the annealing process, marked with the symbols $\vartriangle$
and $\star$ in both figures. For QA, we show the results for 15 instances
with $N=4001$, $y=256$, $\tau=4$, using the mode of the replicas
$\sigma_{j}^{\star}=\mathrm{sgn}\left(\sum_{a=1}^{y}\sigma_{j}^{a}\right)$
as the reference point; for SA, we show 15 samples for $N=4001$ and
$\tau=16$. These results show a marked qualitative difference in
the type of landscape that is typically explored by the two algorithms:
the local landscape of QA is generally much wider, while SA is typically
working inside narrow regions of the landscape which tend to trap
the algorithm eventually. Panel \textbf{c}: local entropy, i.e. the
logarithm of the number of solutions surrounding the reference point
at a given distance $d$ for the same configurations of panel \textbf{a}.
The QA configurations (orange curves at the top) are located in regions
with exponentially many solutions surrounding them (although these
regions are not maximally dense, as can be seen from the comparison
with the dashed curve representing the overall number of surrounding
configurations at that distance). The SA configurations (gray curves
at the bottom) are far away from these exponentially dense regions
(the local entropy has a gap around $d=0$).}
\end{figure*}

In Fig.~\ref{fig:en_landscapes} we provide the profiles of the the
classical energy minima found for different values of $\Gamma$ in
the case of SQA and different temperatures for SA. These results are
computed analytically by the cavity method (see Materials and Methods
and SI for details) by evaluating which is the most probable energy
found at a normalized Hamming distance $d$ from a given configuration.
As it turns out, throughout the annealing process, SQA follows a path
corresponding to wide valleys while SA gets stuck in steep metastable
states. The quantum fluctuations reproduced by the SQA process drive
the system to converge toward wide flat regions, in spite of the fact
that they are exponentially rare compared to the narrow minima.

The physical interpretation of these results is that quantum fluctuations
lower the energy of a cluster proportionally to its size or, in other
words, that quantum fluctuations allow the system to lower its kinetic
energy by delocalizing, see Refs.~\cite{foini2010solvable,markland2011,biroli2012tentative}
for related results. Along the process of reduction of the transverse
field we do not observe any phase transition which could induce a
critical slowing down of the quantum annealing process and we expect
SQA and QA to behave similarly \cite{bapst2013thermal,bapst2013quantum}.

This is in agreement with the results of a direct comparison between
the real time quantum dynamics and the SQA on small systems ($N=21$):
as reported in the Appendix Sec.~\ref{sec:Real-time}, we have performed
extensive numerical studies of properly selected small instances of
the binary perceptron problem, comparing the results of SQA and QA
and analyzing the results of the QA process and the properties of
the Hamiltonian. To reproduce the conditions that are known to exist
at large values of $N$, we have selected instances for which a fast
annealing schedule SA gets trapped at some positive fraction of violated
constraints, and yet the problems display a sufficiently high number
of solutions. We found that the agreement between SQA and QA on each
sample is excellent. The measurements on the final configurations
reached by QA qualitatively confirm the scenario described above,
that QA is attracted towards dense low-energy regions without getting
stuck during the annealing process. Finally, the analysis of the gap
between the ground state of the system and the first excited state
as $\Gamma$ decreases shows no signs of the kind of phenomena which
would typically hamper the performance of QA in other models: there
are no vanishingly small gaps at finite $\Gamma$ (cf.~the discussion
in the introduction). We benchmarked all these results with ``randomized''
versions of the same samples, in which we randomly permuted the classical
energies associated to each spin configuration, so as to keep the
distribution of the classical energy levels while destroying the geometric
structure of the states. Indeed, for these randomized samples, we
found that the gaps nearly close at finite $\Gamma\simeq0.4$, and
that correspondingly the QA process fails to track the ground state
of the system, resulting in a much reduced probability of finding
a solution to the problem.

To reproduce the conditions that are known to exist at large values
of $N$ we have selected instances for which a fast annealing schedule
SA gets trapped at some positive fraction of violated constraints
and yet the problems display a sufficiently high number of solutions.
We have then compared the behavior of SQA and the real time quantum
dynamics studied by the Lanczos method as discussed in \cite{schneider2016chapter}.
The agreement between SQA and QA is ... almost perfect.

As concluding remarks we report that the models with $n=0$ and $n=1$
have phase diagrams which are qualitatively very similar (for the
sake of simplicity, here we reported the $n=0$ case only). The former
presents at very small positive values of $\Gamma$ a collapse of
the density matrix onto the classical one whereas the latter ends
up in the classical state only at $\Gamma=0$.

For the sake of completeness, we have checked that the performance
of SQA in the $y\to\infty$ quantum limit extends to more complex
architectures which include hidden layers; the details are reported
in the Appendix Sec.~\ref{subsec:committee}.

\section{Conclusions}

We conclude by noticing that, at variance with other studies on spin
glass models in which the evidence for QA outperforming classical
annealing was limited to finite values of $y$, thereby just defining
a different type of classical SA algorithms, in our case the quantum
limit coincides with the optimal behavior of the algorithm itself.
We believe that these results could play a role in many optimization
problems in which optimality of the cost function needs to also meet
robustness conditions (i.e.~wide minima). As far as learning problems
are concerned, it is worth mentioning that for the best performing
artificial neural networks, the so called deep networks \cite{lecun2015deep},
there is numerical evidence for the existence of rare flat minima
\cite{keskar2016large}, and that all the effective algorithms always
include effects of systematic injected noise in the learning phase
\cite{bottou2016optimization}, which implies that the equilibrium
Gibbs measure is not the stationary measure of the learning protocols.
For the sake of clarity we should remark that our results are aimed
to suggest that QA can equilibrate efficiently whereas SA cannot,
i.e.~our notion of quantum speed up is relative to the same algorithmic
scheme that runs on classical hardware. Other classical algorithms
for the same class of problems, besides the above-mentioned ones based
on the RE and the SQA itself, have been discovered \cite{braunstein-zecchina,baldassi-et-all-pnas,baldassi-2009,baldassi2015max,hubara2016quantized};
however, all of these algorithms are qualitatively different from
QA, which can provide a huge speed up by manipulating single bits
in parallel. Thus, the overall solving time in a physical QA implementation
(neglecting any other technological considerations) would have, at
worst, only a mild dependence on $N$.

Our results provide further evidence that learning can be achieved
through different types of correlated fluctuations, among which quantum
tunneling could be a relevant example for physical devices.
\begin{acknowledgments}
The authors thank G. Santoro, B. Kappen and F. Becca for discussions.
\end{acknowledgments}

\appendix

\section{Theoretical analysis by the replica method\label{sec:analysis}}

We present here the analytical calculations performed to derive all
the theoretical results mentioned in the main text. For completeness,
we report all the relevant formulas and definitions here, even those
that were already introduced in the main text.

The Hamiltonian operator of a model of $N$ quantum spins with an
energy term acting in the longitudinal direction $z$ and a magnetic
field $\Gamma$ acting in the transverse direction $x$ is written
as:
\begin{equation}
\hat{H}=E\left(\left\{ \hat{\sigma}_{j}^{z}\right\} _{j}\right)-\Gamma\sum_{j=1}^{N}\hat{\sigma}_{j}^{x}
\end{equation}
where $\hat{\sigma}_{j}^{z}$ and $\hat{\sigma_{j}}^{x}$ are the
spin operators (Pauli matrices) in the $z$ and $x$ directions. We
want to study the partition function:
\begin{equation}
Z=\mathrm{Tr}\,\left(e^{-\beta\hat{H}}\right).
\end{equation}

By using the Suzuki-Trotter transformation, we end up with a classical
effective Hamiltonian acting on a system of $y$ interacting Trotter
replicas, to be studied in the limit $y\to\infty$:
\begin{equation}
H_{\mathrm{eff}}\left(\left\{ \sigma_{j}^{a}\right\} _{j,a}\right)=\frac{1}{y}\sum_{a}E\left(\left\{ \sigma_{j}^{a}\right\} _{j}\right)-\frac{\gamma}{\beta}\sum_{aj}\sigma_{j}^{a}\sigma_{j}^{a+1}-\frac{NK}{\beta}
\end{equation}
where the $\sigma_{j}^{a}=\pm1$ are Ising spins, $a\in\left\{ 1,\dots,y\right\} $
is a replica index with periodic boundary conditions $\sigma_{j}^{y+1}\equiv\sigma_{j}^{1}$,
and we have defined:
\begin{eqnarray}
\gamma & = & \frac{1}{2}\log\coth\left(\frac{\beta\Gamma}{y}\right),\\
K & = & \frac{1}{2}y\log\left(\frac{1}{2}\sinh\left(2\frac{\beta\Gamma}{y}\right)\right).
\end{eqnarray}

In the following, we will just use $\sigma^{a}$ to denote the configuration
of one Trotter replica, $\left\{ \sigma_{j}^{a}\right\} _{j}$; we
will always use the indices $a$ or $b$ for the Trotter replicas
and assume that they range in $1,\dots,y$; we will also use $j$
for the site index and assume that it ranges in $1,\dots,N$.

The effective partition function for a given $y$ reads:
\begin{equation}
Z_{\mathrm{eff}}=\sum_{\left\{ \sigma^{a}\right\} }e^{-\frac{\beta}{y}\sum_{a}E\left(\sigma^{a}\right)+\gamma\sum_{aj}\sigma_{j}^{a}\sigma_{j}^{a+1}+NK}.\label{eq:Zeff}
\end{equation}

Here, we first study the binary perceptron case in which the longitudinal
energy $E$ is defined in terms of a set of $\alpha N$ patterns $\left\{ \xi^{\mu}\right\} _{\mu}$
with $\mu\in\left\{ 1,\dots,\alpha N\right\} $, where each pattern
is a binary vector of length $N$, $\xi_{j}^{\mu}=\pm1$:
\begin{equation}
E\left(\sigma\right)=\sum_{\mu=1}^{\alpha N}\Theta\left(-\frac{1}{\sqrt{N}}\sum_{j}\xi_{j}^{\mu}\sigma_{j}\right)\label{eq:E-1}
\end{equation}
where $\Theta\left(x\right)$ is the Heaviside step function: $\Theta\left(x\right)=1$
if $x>0$, $\Theta\left(x\right)=0$ otherwise. The energy thus simply
counts the number of classification errors of the perceptron, assuming
that the desired output for each pattern in the set is $1$ (this
choice can always be made without loss of generality for this model
when the input patterns are random i.i.d. as described below). A different
form for the energy function is treated in sec.~\ref{subsec:E-stability}.

We consider the case in which the patterns entries are extracted randomly
and independently from an unbiased distribution, $P\left(\xi_{j}^{\mu}\right)=\frac{1}{2}\delta\left(\xi_{j}^{\mu}-1\right)+\frac{1}{2}\delta\left(\xi_{j}^{\mu}+1\right)$,
and we want to study the typical properties of this system by averaging
over the quenched disorder introduced by the patterns. We use the
replica method, which exploits the transformation:
\begin{eqnarray}
\left\langle \log Z\right\rangle _{\xi} & = & \lim_{n\to0}\frac{\left\langle Z^{n}\right\rangle _{\xi}-1}{n}\nonumber \\
 & = & \lim_{n\to0}\frac{\left\langle \prod_{c=1}^{n}Z\right\rangle _{\xi}-1}{n}
\end{eqnarray}
where $\left\langle \cdot\right\rangle _{\xi}$ denotes the average
over the disorder. We thus need to replicate the whole system $n$
times, and therefore we have two replica indices for each spin. We
will use indices $c,d=1,\dots,n$ for the ``virtual'' replicas introduced
by the replica method,\footnote{Note that the parameter $n$ has a different meaning in main text,
cf.~sec.~\ref{subsec:E-stability}.} to distinguish them from the indices $a$ and $b$ used for the Trotter
replicas. The average replicated partition function of eq.~(\ref{eq:Zeff})
is thus written as:
\begin{equation}
\left\langle Z_{\mathrm{eff}}^{n}\right\rangle _{\xi}=e^{nNK}\left\langle \int\prod_{caj}d\mu\left(\sigma_{j}^{ca}\right)\prod_{caj}e^{\gamma\sigma_{j}^{c}\sigma_{j}^{c\left(a+1\right)}}\prod_{\mu ca}\left(\Theta\left[\frac{1}{\sqrt{N}}\sum_{j}\xi_{j}^{\mu}\sigma_{j}^{ca}\right]\left(1-e^{-\frac{\beta}{y}}\right)+e^{-\frac{\beta}{y}}\right)\right\rangle _{\xi}
\end{equation}
where we changed the sum over all configurations into an ($n\times y\times N$-dimensional)
integral, using the customary notation $d\mu\left(\sigma\right)=\delta\left(\sigma-1\right)+\delta\left(\sigma+1\right)$
with $\delta\left(\cdot\right)$ denoting the Dirac-delta distribution.
Here and in the following, all integrals will be assumed to range
over the whole $\mathbb{R}$ unless otherwise specified.

We introduce new auxiliary variables $\lambda_{\mu}^{ca}=\frac{1}{\sqrt{N}}\sum_{j}\xi_{j}^{\mu}\sigma_{j}^{ca}$
via additional Dirac-deltas:

\begin{eqnarray}
\left\langle Z_{\mathrm{eff}}^{n}\right\rangle _{\xi} & = & e^{nNK}\int\prod_{caj}d\mu\left(\sigma_{j}^{ca}\right)\prod_{caj}e^{\gamma\sigma_{j}^{ca}\sigma_{j}^{c\left(a+1\right)}}\int\prod_{\mu ca}d\lambda_{\mu}^{ca}\prod_{\mu ca}\left(\Theta\left[\lambda_{\mu}^{ca}\right]\left(1-e^{-\frac{\beta}{y}}\right)+e^{-\frac{\beta}{y}}\right)\times\nonumber \\
 &  & \hfill\qquad\quad\times\left\langle \prod_{\mu ca}\delta\left(\lambda_{\mu}^{ca}-\frac{1}{\sqrt{N}}\sum_{j}\xi_{j}^{\mu}\sigma_{j}^{ca}\right)\right\rangle _{\xi}
\end{eqnarray}

We then use the integral representation of the delta $\delta\left(x\right)=\int\frac{d\hat{x}}{2\pi}e^{ix\hat{x}}$,
and perform the average over the disorder, to the leading order in
$N$:
\begin{eqnarray}
 &  & \left\langle \prod_{\mu ca}\delta\left(\lambda_{\mu}^{ca}-\frac{1}{\sqrt{N}}\sum_{j}\xi_{j}^{\mu}\sigma_{j}^{ca}\right)\right\rangle _{\xi}=\nonumber \\
 &  & \int\prod_{\mu ca}\frac{d\hat{\lambda}_{\mu}^{ca}}{2\pi}\prod_{\mu ca}e^{i\hat{\lambda}_{\mu}^{ca}\lambda_{\mu}^{ca}}\prod_{\mu}\exp\left(-\frac{1}{2}\sum_{cdab}\hat{\lambda}_{\mu}^{ca}\hat{\lambda}_{\mu}^{db}\left(\frac{1}{N}\sum_{j}\sigma_{j}^{ca}\sigma_{j}^{db}\right)\right)
\end{eqnarray}

Next, we introduce the overlaps $q^{ca,db}=\frac{1}{N}\sum_{j}\sigma_{j}^{ca}\sigma_{j}^{db}$
via Dirac-deltas (note that due to symmetries and the fact that the
self-overlaps are always $1$ we have $ny\left(ny-1\right)/2$ overlaps
overall), expand those deltas introducing conjugate parameters $\hat{q}^{ca,db}$
(as usual for these parameters in these models, we absorb away a factor
$i$ and integrate them along the imaginary axis, without explicitly
noting this), and finally factorize over the site and pattern indices:
\begin{eqnarray}
\left\langle Z_{\mathrm{eff}}^{n}\right\rangle _{\xi} & = & e^{nNK}\int\prod_{c,a>b}\frac{dq^{ca,cb}d\hat{q}^{ca,cb}N}{2\pi}\prod_{c>d,ab}\frac{dq^{ca,db}d\hat{q}^{ca,db}N}{2\pi}\times\nonumber \\
 &  & \quad\times e^{-N\sum_{c,a>b}q^{ca,cb}\hat{q}^{ca,cb}-N\sum_{c>d,ab}q^{ca,db}\hat{q}^{ca,db}}\times G_{S}^{N}\times G_{E}^{\alpha N}\\
G_{S} & \doteq & \int\prod_{ca}d\mu\left(\sigma^{ca}\right)e^{\sum_{c,a>b}\hat{q}^{ca,cb}\sigma^{ca}\sigma^{cb}+\sum_{c>d,ab}\hat{q}^{ca,db}\sigma^{ca}\sigma^{db}+\gamma\sum_{ca}\sigma^{ca}\sigma^{c\left(a+1\right)}}\\
G_{E} & \doteq & \int\prod_{ca}\frac{d\lambda^{ca}d\hat{\lambda}^{ca}}{2\pi}\prod_{ca}\left(\Theta\left[\lambda^{ca}\right]\left(1-e^{-\frac{\beta}{y}}\right)+e^{-\frac{\beta}{y}}\right)\times\\
 &  & \quad\times e^{-\frac{1}{2}\sum_{ca}\left(\hat{\lambda}^{ca}\right)^{2}+i\sum_{ca}\lambda^{ca}\hat{\lambda}^{ca}-\sum_{c,a>b}\hat{\lambda}^{ca}\hat{\lambda}^{cb}q^{ca,cb}-\sum_{c>d,ab}\hat{\lambda}^{ca}\hat{\lambda}^{db}q^{ca,db}}\nonumber 
\end{eqnarray}

We now introduce the replica-symmetric (RS) ansatz for the overlaps:
\begin{equation}
q^{ca,db}=\begin{cases}
q_{1} & \mathrm{if}\:c=d\\
q_{0} & \mathrm{if}\:c\ne d
\end{cases}
\end{equation}
and analogous for the conjugate parameters $\hat{q}^{ca,db}$.

Note that this is the so-called ``static approximation'' since we
neglect the dependency of the overlap from the distance along the
Trotter dimension; however, we have kept the interaction term $\gamma\sum_{ca}\sigma^{ca}\sigma^{c\left(a+1\right)}$
and inserted it in the $G_{S}$ term (rather than writing it in terms
of the overlap $q^{ca,c\left(a+1\right)}$ and inserting it in the
$G_{E}$ term where it would have been rewritten as $\gamma q_{1}$).
This difference, despite its inconsistency, is the standard procedure
when performing the static approximation, and is justified a posteriori
from the comparison with the numerical simulation results. We obtain:
\begin{eqnarray}
\left\langle Z_{\mathrm{eff}}^{n}\right\rangle _{\xi} & = & e^{nNK}\int\prod_{c,a>b}\frac{dq^{ca,cb}d\hat{q}^{ca,cb}N}{2\pi}\prod_{c>d,ab}\frac{dq^{ca,db}d\hat{q}^{ca,db}N}{2\pi}\times\nonumber \\
 &  & \quad\times e^{-Nn\frac{y\left(y-1\right)}{2}q_{1}\hat{q}_{1}-N\frac{n\left(n-1\right)}{2}y^{2}q_{0}\hat{q}_{0}}\times G_{S}^{N}\times G_{E}^{\alpha N}\\
G_{S} & = & \int\prod_{ca}d\mu\left(\sigma^{ca}\right)e^{\hat{q}_{1}\sum_{c,a>b}\sigma^{ca}\sigma^{cb}+\hat{q}_{0}\sum_{c>d,ab}\sigma^{ca}\sigma^{db}+\gamma\sum_{ca}\sigma^{ca}\sigma^{c\left(a+1\right)}}\\
G_{E} & = & \int\prod_{ca}\frac{d\lambda^{ca}d\hat{\lambda}^{ca}}{2\pi}\prod_{ca}\left(\Theta\left[\lambda^{ca}\right]\left(1-e^{-\frac{\beta}{y}}\right)+e^{-\frac{\beta}{y}}\right)\times\\
 &  & \quad\times e^{-\frac{1}{2}\sum_{ca}\left(\hat{\lambda}^{ca}\right)^{2}+i\sum_{ca}\lambda^{ca}\hat{\lambda}^{ca}-q_{1}\sum_{c,a>b}\hat{\lambda}^{ca}\hat{\lambda}^{cb}-q_{0}\sum_{c>d,ab}\hat{\lambda}^{ca}\hat{\lambda}^{db}}\nonumber 
\end{eqnarray}

The entropic term $G_{S}$ can be explicitly computed as
\begin{eqnarray}
G_{S} & = & \int\prod_{ca}d\mu\left(\sigma^{ca}\right)e^{\frac{1}{2}\hat{q}_{1}\sum_{c}\left(\left(\sum_{a}\sigma^{ca}\right)^{2}-\sum_{a}\left(\sigma^{ca}\right)^{2}\right)+\frac{1}{2}\hat{q}_{0}\left(\left(\sum_{ca}\sigma^{ca}\right)^{2}-\sum_{c}\left(\sum_{a}\sigma^{ca}\right)^{2}\right)}\nonumber \\
 &  & \quad\times e^{\gamma\sum_{ca}\sigma^{ca}\sigma^{c\left(a+1\right)}}\nonumber \\
 & = & e^{-\frac{1}{2}\hat{q}_{1}ny}\int\prod_{ca}d\mu\left(\sigma^{ca}\right)e^{\frac{1}{2}\left(\hat{q}_{1}-\hat{q}_{0}\right)\sum_{c}\left(\sum_{a}\sigma^{ca}\right)^{2}+\frac{1}{2}\hat{q}_{0}\left(\sum_{ca}\sigma^{ca}\right)^{2}+\gamma\sum_{ca}\sigma^{ca}\sigma^{c\left(a+1\right)}}\nonumber \\
 & = & \int Dz_{0}\,e^{-\frac{1}{2}\hat{q}_{1}ny}\left[\int\prod_{a}d\mu\left(\sigma^{a}\right)e^{\frac{1}{2}\left(\hat{q}_{1}-\hat{q}_{0}\right)\left(\sum_{a}\sigma^{a}\right)^{2}+z_{0}\sqrt{\hat{q}_{0}}\left(\sum_{a}\sigma^{a}\right)+\gamma\sum_{a}\sigma^{a}\sigma^{a+1}}\right]^{n}\nonumber \\
 & = & \int Dz_{0}\,e^{-\frac{1}{2}\hat{q}_{1}ny}\left[\int Dz_{1}\int\prod_{a}d\mu\left(\sigma^{a}\right)e^{\left(z_{1}\sqrt{\hat{q}_{1}-\hat{q}_{0}}+z_{0}\sqrt{\hat{q}_{0}}\right)\left(\sum_{a}\sigma^{a}\right)+\gamma\sum_{a}\sigma^{a}\sigma^{a+1}}\right]^{n}
\end{eqnarray}
where the notation $Dz=dz\,\frac{1}{\sqrt{2\pi}}e^{-\frac{x^{2}}{2}}$
is a shorthand for a Gaussian integral, and we used twice the Hubbard-Stratonovich
transformation $e^{\frac{1}{2}b}=\int Dz\,e^{z\sqrt{b}}$. The expression
between square brackets in the last line is the partition function
of a 1-dimensional Ising model of size $y$ with uniform interactions
$J=\gamma$ and uniform fields $h=z_{1}\sqrt{\hat{q}_{1}-\hat{q}_{0}}+z_{0}\sqrt{\hat{q}_{0}}$
and can be computed by the well-known transfer matrix method. Note
however that while usually in the analysis of the 1D Ising spin model
it is sufficient to keep the largest eigenvalue of the transfer matrix
in the thermodynamic limit $y\to\infty$, in this case instead we
need to keep both eigenvalues, since the interaction term scales with
the size of the system. The result is:
\begin{eqnarray}
G_{S} & = & \int Dz_{0}\,e^{-\frac{1}{2}\hat{q}_{1}ny}\left[\int Dz_{1}e^{\gamma y}\sum_{w=\pm1}g\left(z_{0},z_{1},w\right)^{y}\right]^{n}\\
g\left(z_{0,}z_{1},w\right) & \doteq & \cosh\left(h\left(z_{0},z_{1}\right)\right)+w\sqrt{\sinh\left(h\left(z_{0},z_{1}\right)\right)^{2}+e^{-4\gamma}}\\
h\left(z_{0},z_{1}\right) & \doteq & z_{1}\sqrt{\hat{q}_{1}-\hat{q}_{0}}+z_{0}\sqrt{\hat{q}_{0}}
\end{eqnarray}

In the limit of small $n$ we obtain:
\begin{eqnarray}
\mathcal{G}_{S} & \doteq & \frac{1}{n}\log G_{S}+\frac{1}{2}\hat{q}_{1}y-\gamma y\nonumber \\
 & = & \int Dz_{0}\,\log\left[\int Dz_{1}\sum_{w=\pm1}\left(\cosh\left(h\left(z_{0},z_{1}\right)\right)+w\sqrt{\sinh\left(h\left(z_{0},z_{1}\right)\right)^{2}+e^{-4\gamma}}\right)^{y}\right]\label{eq:Gs_finite_y}
\end{eqnarray}

Note that in the limit of large $y$ the term $\gamma y$ tends to$-K$
up to terms of order $y^{-1}$.

The energetic term $G_{E}$ is computed similarly, by first performing
two Hubbard-Stratonovich transformations which allow to factorize
the indices $c$ and $a$, and then explicitly performing the inner
integrals:
\begin{eqnarray}
G_{E} & = & \int\prod_{ca}\frac{d\lambda^{ca}d\hat{\lambda}^{ca}}{2\pi}\prod_{ca}\left(\Theta\left[\lambda^{ca}\right]\left(1-e^{-\frac{\beta}{y}}\right)+e^{-\frac{\beta}{y}}\right)\times\nonumber \\
 &  & \quad\times e^{-\frac{1}{2}\sum_{ca}\left(\hat{\lambda}^{ca}\right)^{2}+i\sum_{ca}\lambda^{ca}\hat{\lambda}^{ca}-\frac{1}{2}q_{1}\sum_{c}\left(\left(\sum_{a}\hat{\lambda}^{ca}\right)^{2}-\sum_{a}\left(\hat{\lambda}^{ca}\right)^{2}\right)-\frac{1}{2}q_{0}\left(\left(\sum_{ca}\hat{\lambda}^{ca}\right)^{2}-\sum_{c}\left(\sum_{a}\hat{\lambda}^{ca}\right)^{2}\right)}\nonumber \\
 & = & \int Dz_{0}\left[\int Dz_{1}\left[\int\frac{d\lambda d\hat{\lambda}}{2\pi}\left(\Theta\left[\lambda\right]\left(1-e^{-\frac{\beta}{y}}\right)+e^{-\frac{\beta}{y}}\right)e^{-\frac{1-q_{1}}{2}\left(\hat{\lambda}\right)^{2}+i\hat{\lambda}\left(\lambda-z_{1}\sqrt{q_{1}-q_{0}}-z_{0}\sqrt{q_{0}}\right)}\right]^{y}\right]^{n}\nonumber \\
 & = & \int Dz_{0}\left[\int Dz_{1}\left[1-\left(1-e^{-\frac{\beta}{y}}\right)H\left(\frac{z_{1}\sqrt{q_{1}-q_{0}}+z_{0}\sqrt{q_{0}}}{\sqrt{1-q_{1}}}\right)\right]^{y}\right]^{n}\label{eq:Ge_finite_yn}
\end{eqnarray}
where $H\left(x\right)=\frac{1}{2}\mathrm{erfc}\left(\frac{x}{\sqrt{2}}\right)$.
In the limit of small $n$ and of large $y$ we finally obtain:
\begin{equation}
\mathcal{G}_{E}\doteq\frac{1}{n}\log G_{E}=\int Dz_{0}\log\int Dz_{1}\exp\left(-\beta H\left(\frac{z_{1}\sqrt{q_{1}-q_{0}}+z_{0}\sqrt{q_{0}}}{\sqrt{1-q_{1}}}\right)\right)\label{eq:Ge_final-pre}
\end{equation}

Using equations~(\ref{eq:Gs_finite_y}) and~(\ref{eq:Ge_final-pre}),
we obtain the expression for the action:
\begin{equation}
\phi\doteq\frac{1}{N}\left\langle \log Z_{\mathrm{eff}}\right\rangle =\mathrm{extr}_{q_{0},q_{1},\hat{q}_{0},\hat{q}_{1}}\left\{ \frac{1}{2}y^{2}q_{0}\hat{q}_{0}-\frac{1}{2}y\left(y-1\right)q_{1}\hat{q}_{1}-\frac{1}{2}\hat{q}_{1}y+\mathcal{G}_{S}+\alpha\mathcal{G}_{E}\right\} 
\end{equation}

In order to obtain a finite result in the limit of $y\to\infty$,
we assume the following scalings for the conjugated order parameters:
\begin{eqnarray}
\hat{q}_{0} & = & \frac{\hat{p}_{0}}{y^{2}}\\
\hat{q}_{1} & = & \frac{\hat{p}_{1}}{y^{2}}
\end{eqnarray}

With these, we find the following final expressions:
\begin{eqnarray}
\phi & = & \mathrm{extr}_{q_{0},q_{1},\hat{p}_{0},\hat{p}_{1}}\left\{ \frac{1}{2}q_{0}\hat{p}_{0}-\frac{1}{2}q_{1}\hat{p}_{1}+\mathcal{G}_{S}+\alpha\mathcal{G}_{E}\right\} \label{eq:phi_final}\\
\mathcal{G}_{S} & = & \int Dz_{0}\,\log\left[\int Dz_{1}\,2\cosh\left(\sqrt{\hat{k}\left(z_{0},z_{1}\right)^{2}+\beta^{2}\Gamma^{2}}\right)\right]\label{eq:Gs_final}\\
\hat{k}\left(z_{0},z_{1}\right) & = & z_{1}\sqrt{\hat{p}_{1}-\hat{p}_{0}}+z_{0}\sqrt{\hat{p}_{0}}\label{eq:k_hat}\\
\mathcal{G}_{E} & = & \int Dz_{0}\log\int Dz_{1}\exp\left(-\beta H\left(k\left(z_{0},z_{1}\right)\right)\right)\label{eq:Ge_final}\\
k\left(z_{0},z_{1}\right) & = & \frac{z_{1}\sqrt{q_{1}-q_{0}}+z_{0}\sqrt{q_{0}}}{\sqrt{1-q_{1}}}\label{eq:k}
\end{eqnarray}

The parameters $q_{0}$, $q_{1}$, $\hat{p}_{0}$ and $\hat{p}_{1}$
are found by solving the system of equations obtained by setting the
partial derivatives of $\phi$ with respect to those parameters to
$0$:
\begin{eqnarray}
\hat{p}_{0} & = & \frac{\alpha\beta}{\sqrt{1-q1}}\int Dz_{0}\frac{\int Dz_{1}e^{-\beta H\left(k\left(z_{0},z_{1}\right)\right)}G\left(k\left(z_{0},z_{1}\right)\right)\left(\frac{z_{1}}{\sqrt{q_{1}-q_{0}}}-\frac{z_{0}}{\sqrt{q_{0}}}\right)}{\int Dz_{1}e^{-\beta H\left(k\left(z_{0},z_{1}\right)\right)}}\label{eq:p0_hat}\\
\hat{p}_{1} & = & \frac{\alpha\beta}{\sqrt{\left(1-q_{1}\right)^{3}\left(q_{1}-q_{0}\right)}}\times\label{eq:p1_hat}\\
 &  & \times\int Dz_{0}\frac{\int Dz_{1}e^{-\beta H\left(k\left(z_{0},z_{1}\right)\right)}G\left(k\left(z_{0},z_{1}\right)\right)\left(z_{0}\sqrt{q_{0}\left(q_{1}-q_{0}\right)}+z_{1}\left(1-q_{0}\right)\right)}{\int Dz_{1}e^{-\beta H\left(k\left(z_{0},z_{1}\right)\right)}}\nonumber \\
q_{0} & = & \frac{1}{\sqrt{\hat{k}\left(z_{0},z_{1}\right)^{2}+\beta^{2}\Gamma^{2}}}\times\\
 &  & \times\int Dz_{0}\frac{\int Dz_{1}\sinh\left(\sqrt{\hat{k}\left(z_{0},z_{1}\right)^{2}+\beta^{2}\Gamma^{2}}\right)\hat{k}\left(z_{0},z_{1}\right)\left(\frac{z_{1}}{\sqrt{\hat{p}_{1}-\hat{p}_{0}}}-\frac{z_{0}}{\sqrt{\hat{p}_{0}}}\right)}{\int Dz_{1}\cosh\left(\sqrt{\hat{k}\left(z_{0},z_{1}\right)^{2}+\beta^{2}\Gamma^{2}}\right)}\nonumber \\
q_{1} & = & \frac{1}{\sqrt{\hat{k}\left(z_{0},z_{1}\right)^{2}+\beta^{2}\Gamma^{2}}}\times\label{eq:q1}\\
 &  & \times\int Dz_{0}\frac{\int Dz_{1}\sinh\left(\sqrt{\hat{k}\left(z_{0},z_{1}\right)^{2}+\beta^{2}\Gamma^{2}}\right)\hat{k}\left(z_{0},z_{1}\right)\left(\frac{z_{1}}{\sqrt{\hat{p}_{1}-\hat{p}_{0}}}\right)}{\int Dz_{1}\cosh\left(\sqrt{\hat{k}\left(z_{0},z_{1}\right)^{2}+\beta^{2}\Gamma^{2}}\right)}\nonumber 
\end{eqnarray}

Once these are found, we can use them to compute the action $\phi$
and the average values of the longitudinal energy and the transverse
fields, and finally of the Hamiltonian:
\begin{eqnarray}
\overline{\left\langle \hat{H}\right\rangle _{\xi}} & = & N\left(\bar{E}-\Gamma\bar{T}\right)\label{eq:H-QA}\\
\bar{E} & = & \frac{1}{N}\overline{\left\langle E\left(\left\{ \hat{\sigma}^{z}\right\} \right)\right\rangle _{\xi}}=-\frac{\partial\phi}{\partial\beta}=\alpha\int Dz_{0}\frac{\int Dz_{1}e^{-\beta H\left(k\left(z_{0},z_{1}\right)\right)}H\left(k\left(z_{0},z_{1}\right)\right)}{\int Dz_{1}e^{-\beta H\left(k\left(z_{0},z_{1}\right)\right)}}\label{eq:E-QA}\\
\bar{T} & = & \frac{1}{N}\overline{\left\langle \hat{\sigma}_{j}^{x}\right\rangle }=\frac{\partial\phi}{\partial\left(\beta\Gamma\right)}=\int Dz_{0}\frac{\int Dz_{1}\frac{\beta\Gamma\,\sinh\left(\sqrt{\hat{k}\left(z_{0},z_{1}\right)^{2}+\beta^{2}\Gamma^{2}}\right)}{\sqrt{\hat{k}\left(z_{0},z_{1}\right)^{2}+\beta^{2}\Gamma^{2}}}}{\int Dz_{1}\cosh\left(\sqrt{\hat{k}\left(z_{0},z_{1}\right)^{2}+\beta^{2}\Gamma^{2}}\right)}
\end{eqnarray}
where the notation $\overline{\left\langle \cdot\right\rangle _{\xi}}$
denotes the fact that we performed both the average over the quenched
disorder and the thermal average.

\subsubsection{Small $\Gamma$ limit\label{subsec:small-Gamma-n0}}

It can be verified that in the limit $\Gamma\to0$ the equations~(\ref{eq:phi_final})-(\ref{eq:k})
reduce to the classical case, in the RS description. In this limit,
$q_{1}\to1$ (i.e., the Trotter replicas collapse), which leads to:
\begin{equation}
\mathcal{G}_{E}=\int Dz_{0}\log\left(\left(1-e^{-\beta}\right)H\left(z_{0}\sqrt{\frac{q_{0}}{1-q_{0}}}\right)+e^{-\beta}\right).
\end{equation}

For $\Gamma=0$ and $q_{1}=1$ we also have the identity:\footnote{This follows from $\int Dz_{1}\cosh\left(a\,z_{1}+b\,z_{0}\right)=e^{\frac{a^{2}}{2}}\cosh\left(b\,z_{0}\right)$.}
\begin{equation}
-\frac{1}{2}\hat{p}_{1}q_{1}+\mathcal{G}_{S}=-\frac{1}{2}\hat{p}_{0}+\int Dz_{0}\log2\cosh\left(z_{0}\sqrt{\hat{p}_{0}}\right).\label{eq:magic}
\end{equation}

Putting these two expressions back in eq.~(\ref{eq:phi_final}) we
recover the classical expression where $\hat{p}_{0}$ assumes the
role of the usual conjugate parameter $\hat{q}$ in the RS analysis
of ref.~\cite{krauth-mezard}.

In order to study in detail how this classical limit is reached, however,
we need to expand the saddle point equations around this limit. To
to this, we define $\epsilon=1-q_{1}\ll1$. From equation~(\ref{eq:p1_hat}),
expanding to the leading order, we obtain the scaling $\hat{p}_{1}=\frac{\hat{c}_{1}}{\sqrt{\epsilon}}$,
with
\begin{equation}
\hat{c}_{1}=\left[\frac{1}{\sqrt{1-q_{0}}}\int Dz_{0}\frac{G\left(z_{0}\sqrt{\frac{q_{0}}{1-q_{0}}}\right)}{e^{-\beta}+\left(1-e^{-\beta}\right)H\left(z_{0}\sqrt{\frac{q_{0}}{1-q_{0}}}\right)}\right]\left[\int Dz_{1}\exp\left(-\beta H\left(z_{1}\right)\right)z_{1}\right].\label{eq:c1hat}
\end{equation}
Then, we use this scaling in equation~(\ref{eq:q1}) and we expand
it, first using $\beta\Gamma\ll1$ and then $\epsilon\ll1$. We obtain
the approximate expression:
\begin{equation}
\epsilon=\frac{\beta^{2}\Gamma^{2}}{2}\frac{-\sqrt{\hat{c}_{1}\epsilon}+\sqrt{2}\left(\hat{c}_{1}+\sqrt{\epsilon}\right)\epsilon^{1/4}F\left(\frac{1}{\sqrt{2}}\sqrt{\frac{\hat{c}_{1}}{\sqrt{\epsilon}}}\right)}{\hat{c}_{1}^{3/2}}
\end{equation}
where $F\left(x\right)=\frac{\sqrt{\pi}}{2}e^{-x^{2}}\mathrm{erfi}\left(x\right)$
is the Dawson's function. For a given $\beta$ (from which we obtain
$\hat{c}_{1}$ via eq.~(\ref{eq:c1hat})), this equation can be solved
numerically to obtain $\epsilon$ (and thus $q_{1}$ and $\hat{p}_{1}$)
as a function of $\Gamma$. This expression has always the solution
$\epsilon=0$, which correspond to the purely classical case. There
is a critical $\Gamma$ below which $\epsilon=0$ is also the only
solution; above that, two additional solutions appear at $\epsilon>0$,
of which the largest is the physical one. Therefore, the classical
limit is not achieved continuously, but rather with a first-order
transition (although the step is tiny).

\subsection{Energy function with stability\label{subsec:E-stability}}

We can generalize the energy function eq.~(\ref{eq:E-1}) to take
into account, for those patterns that are misclassified, by how much
the classification is wrong: 
\begin{equation}
E\left(\sigma\right)=\sum_{\mu=1}^{\alpha N}\Theta\left(-\frac{1}{\sqrt{N}}\sum_{j}\xi_{j}^{\mu}\sigma_{j}\right)\left(-\frac{1}{\sqrt{N}}\sum_{j}\xi_{j}^{\mu}\sigma_{j}\right)^{r}.
\end{equation}

The previous case is recovered by setting $r=0$. Here, we study the
case $r=1$. Note that this parameter is called $n$ in the main text:
that notation was borrowed from ref.~\cite{horner1992dynamics},
but here we change it in order to avoid confusion with the number
of replicas. While the ground states in the SAT phase of the classical
model are unaffected, the system can have different properties for
finite $\beta$.

This change only affects the $\mathcal{G}_{E}$ term. Equation~(\ref{eq:Ge_finite_yn})
becomes (with the definition of eq.~(\ref{eq:k})):
\begin{eqnarray}
G_{E} & = & \int Dz_{0}\left[\int Dz_{1}\left[e^{\frac{\beta}{y}\sqrt{1-q_{1}}\left(k\left(z_{0},z_{1}\right)+\frac{1}{2}\frac{\beta}{y}\sqrt{1-q_{1}}\right)}H\left(k\left(z_{0},z_{1}\right)+\frac{\beta}{y}\sqrt{1-q_{1}}\right)+\right.\right.\nonumber \\
 &  & \left.\negmedspace\ \ \ \ \ \ \ \ \ \ \ \ \ \ \ \ \ \ \ \ \ \ +H\left(-k\left(z_{0},z_{1}\right)\right)\Big]^{y}\right]^{n}.
\end{eqnarray}

In the limit of large $y$ we have the modified version of eq.~(\ref{eq:Ge_final-pre}):
\begin{equation}
\mathcal{G}_{E}=\frac{1}{n}\log G_{E}=\int Dz_{0}\log\int Dz_{1}\exp\left(-\beta\sqrt{1-q_{1}}\left[G\left(k\left(z_{0},z_{1}\right)\right)-k\left(z_{0},z_{1}\right)H\left(k\left(z_{0},z_{1}\right)\right)\right]\right)\label{eq:Ge_final_r1}
\end{equation}

The saddle point equations~(\ref{eq:p0_hat}) and~(\ref{eq:p1_hat})
become:
\begin{eqnarray}
\hat{p}_{0} & = & -\alpha\beta\int Dz_{0}\frac{\int Dz_{1}\exp\left(-\beta\sqrt{1-q_{1}}A\left(z_{0},z_{1}\right)\right)H\left(k\left(z_{0},z_{1}\right)\right)\left(\frac{z_{1}}{\sqrt{q_{1}-q_{0}}}-\frac{z_{0}}{\sqrt{q_{0}}}\right)}{\int Dz_{1}\exp\left(\beta\sqrt{1-q_{1}}A\left(z_{0},z_{1}\right)\right)}\label{eq:p0_hat_r1}\\
\hat{p}_{1} & = & \alpha\beta^{2}\int Dz_{0}\frac{\int Dz_{1}\exp\left(-\beta\sqrt{1-q_{1}}A\left(z_{0},z_{1}\right)\right)H\left(k\left(z_{0},z_{1}\right)\right)^{2}}{\int Dz_{1}\exp\left(\beta\sqrt{1-q_{1}}A\left(z_{0},z_{1}\right)\right)}\label{eq:p1_hat_r1}
\end{eqnarray}
where
\[
A\left(z_{0},z_{1}\right)=G\left(k\left(z_{0},z_{1}\right)\right)-k\left(z_{0},z_{1}\right)H\left(k\left(z_{0},z_{1}\right)\right).
\]

\subsubsection{Small $\Gamma$ limit}

As in the previous case, it can be checked that for $\Gamma\to0$,
we have $q_{1}\to1$ and eq.~(\ref{eq:Ge_final_r1}) becomes the
expression for the classical model under the RS ansatz:
\begin{equation}
\mathcal{G}_{E}=\int Dz_{0}\log\left(e^{\beta\sqrt{1-q_{0}}\left(k_{0}\left(z_{0}\right)+\frac{1}{2}\beta\sqrt{1-q_{0}}\right)}H\left(k_{0}\left(z_{0}\right)+\beta\sqrt{1-q_{0}}\right)+H\left(-k_{0}\left(z_{0}\right)\right)\right)
\end{equation}
where $k_{0}\left(z_{0}\right)=z_{0}\sqrt{\frac{q_{0}}{1-q_{0}}}$.
Also, eq.~(\ref{eq:magic}) still holds, and $\hat{p}_{0}$ takes
the role of the usual parameter $\hat{q}$ in the classical RS analysis.
In this case, however, we no longer have $\hat{p}_{1}\to\infty$;
rather, it tends to a finite value:
\begin{equation}
\hat{p}_{1}=\alpha\beta^{2}\int Dz_{0}\left(1-\frac{H\left(-k_{0}\left(z_{0}\right)\right)}{e^{\beta\sqrt{1-q_{0}}\left(k_{0}\left(z_{0}\right)+\frac{1}{2}\beta\sqrt{1-q_{0}}\right)}H\left(k_{0}\left(z_{0}\right)+\beta\sqrt{1-q_{0}}\right)+H\left(-k_{0}\left(z_{0}\right)\right)}\right)
\end{equation}

Therefore, the scaling of $\epsilon=1-q_{1}$ is different in this
case. We find (using the definition of eq.~(\ref{eq:k_hat})):
\begin{equation}
1-q_{1}=\beta^{2}\Gamma^{2}\int Dz_{0}\frac{e^{-\frac{\hat{p}_{1}-\hat{p}_{0}}{2}}}{\cosh\left(z_{0}\sqrt{\hat{p}_{0}}\right)}\int Dz_{1}\frac{1}{\hat{k}\left(z_{0},z_{1}\right)^{2}}\left(\cosh\left(\hat{k}\left(z_{0},z_{1}\right)\right)-\frac{\sinh\left(\hat{k}\left(z_{0},z_{1}\right)\right)}{\hat{k}\left(z_{0},z_{1}\right)}\right)
\end{equation}

Therefore, the convergence to the classical case is smooth.

\section{Estimation of the local energy and entropy landscapes with the cavity
method\label{sec:BP}}

In order to compute the local landscapes of the energy and the entropy
around a reference configuration (Fig.~\ref{fig:en_landscapes}),
we used the Belief Propagation (BP) algorithm, a cavity method message-passing
algorithm that has been successfully employed numerous times for the
study of disordered systems \cite{mezard_information_2009}. In the
case of single-layer binary perceptrons trained on random unbiased
i.i.d. patterns, it is believed that the results of this algorithm
are exact in the limit of $N\to\infty$, at least up to the critical
value $\alpha_{c}\approx0.83$ \cite{mezard_space_1989}.

For a full explanation of the BP equations for binary perceptrons,
we refer the interested reader to the Appendix of ref.~\cite{baldassi_local_2016}.
Here, we provide only a summary. The BP equations involve two sets
of quantities (called ``messages''), representing cavity marginal
probabilities associated with each edge in a factor graph representation
of the (classical) Boltzmann distribution induced by the energy function~(\ref{eq:E-1}).
To each edge in the graph linking the variable node $i$ with the
factor node $\mu$, are associated two messages, $m_{i\to\mu}$ and
$\hat{m}_{\mu\to i}$. These are determined by solving iteratively
the following system of equations:
\begin{eqnarray}
m_{i\to\mu} & = & \tanh\left(\sum_{\nu\neq\mu}\tanh^{-1}\left(\hat{m}_{\nu\to i}\right)\right)\label{eq:bp_perc_var}\\
\hat{m}_{\mu\to i} & = & \xi_{i}\,g\left(a_{\mu\to i},b_{\mu\to i}\right)\label{eq:bp_perc_node}
\end{eqnarray}
where:
\begin{eqnarray}
g\left(a,b\right) & = & \frac{H\left(\frac{a-1}{b}\right)-H\left(\frac{a+1}{b}\right)}{H\left(\frac{a-1}{b}\right)+H\left(\frac{a+1}{b}\right)}\\
a_{\mu\to i} & = & \sum_{j\neq i}\xi_{j}^{\mu}m_{j\to\mu}\\
b_{\mu\to i} & = & \sqrt{\sum_{j\neq i}\left(1-m_{j\to\mu}^{2}\right)}
\end{eqnarray}
(as for the previous section, we used the definition $H\left(x\right)=\frac{1}{2}\mathrm{erfc}\left(\frac{x}{\sqrt{2}}\right)$.)

Once a self-consistent solution is found, these quantities can be
used to compute, using standard formulas, all thermodynamic quantities
of interest, in particular the typical (equilibrium) energy and the
entropy of the system. A numerically accurate implementation of these
equations is available at ref.~\cite{FBP-code}.

It is also possible to compute those same thermodynamic quantities
in a neighborhood of some arbitrary reference configuration $w=\left\{ w_{i}\right\} _{i}$.
This is achieved by adding an external field in the direction of that
configuration, which amounts at this simple modification of eq\@.~(\ref{eq:bp_perc_var}):
\begin{equation}
m_{i\to\mu}=\tanh\left(\sum_{\nu\neq\mu}\tanh^{-1}\left(\hat{m}_{\nu\to i}\right)+\lambda w_{i}\right)
\end{equation}

By varying the auxiliary parameter $\lambda$, we can control the
size of the neighborhood under consideration (the larger $\lambda$,
the narrower the neighborhood); the typical normalized Hamming distance
from the reference of the configurations that are considered by this
modified measure can be obtained from the fixed-point BP messages
for any given $\lambda$ by this formula:
\begin{equation}
d=\frac{1}{2}\left(1-\frac{1}{N}\sum_{i}m_{i}w_{i}\right)
\end{equation}
where the $m_{i}$ are the total magnetizations:
\begin{equation}
m_{i}=\tanh\left(\sum_{\nu}\tanh^{-1}\left(\hat{m}_{\nu\to i}\right)+\lambda w_{i}\right)
\end{equation}

In order to produce the energy landscape plots of Figs.~\ref{fig:en_landscapes}a
and~\ref{fig:en_landscapes}b, we simply ran this algorithm at infinite
temperature, varying $\lambda$ and plotting the energy density shift
from the center as a function of $d$. This gives us an estimate of
the most probable energy density shift which would be obtained by
moving in a random point at distance $d$ from the reference.

The plot in Fig.~\ref{fig:en_landscapes}c was similarly obtained
by setting the temperature to $0$ and computing the entropy density
instead, which in this context is then simply the natural logarithm
of the number of solutions in the given neighborhood, divided by $N$.

\section{Numerical simulations details of the annealing protocols\label{sec:Numerical-simulations-details}}

\subsection{Quantum annealing protocol}

In this section we provide the details of the QA results presented
in Fig.~\ref{fig:sim_vs_theory}. The simulations were performed
using the RRR Monte Carlo method \cite{baldassi_method_2017}. We
fixed the total number of spin flip attempts at $\tau Ny\cdot10^{4}$
and followed a linear protocol for the annealing of $\Gamma$, starting
from $\Gamma_{0}=2.5$ and reaching down $\Gamma_{1}=0$. We actually
divided the annealing in $30\tau$ steps, where during each step $\Gamma$
was kept constant and decreased by $\Delta\Gamma=\frac{\Gamma_{0}-\Gamma_{1}}{30\tau}$
after each step. In the figure, we have shown the results for $N=4001$
and $\tau=4$; the results for $N=1001,2001$ and for $\tau=1,2$
were essentially indistinguishable at that level of detail.

\subsection{Classical simulated annealing protocol}

\begin{figure}
\centering{}\includegraphics[width=0.5\columnwidth]{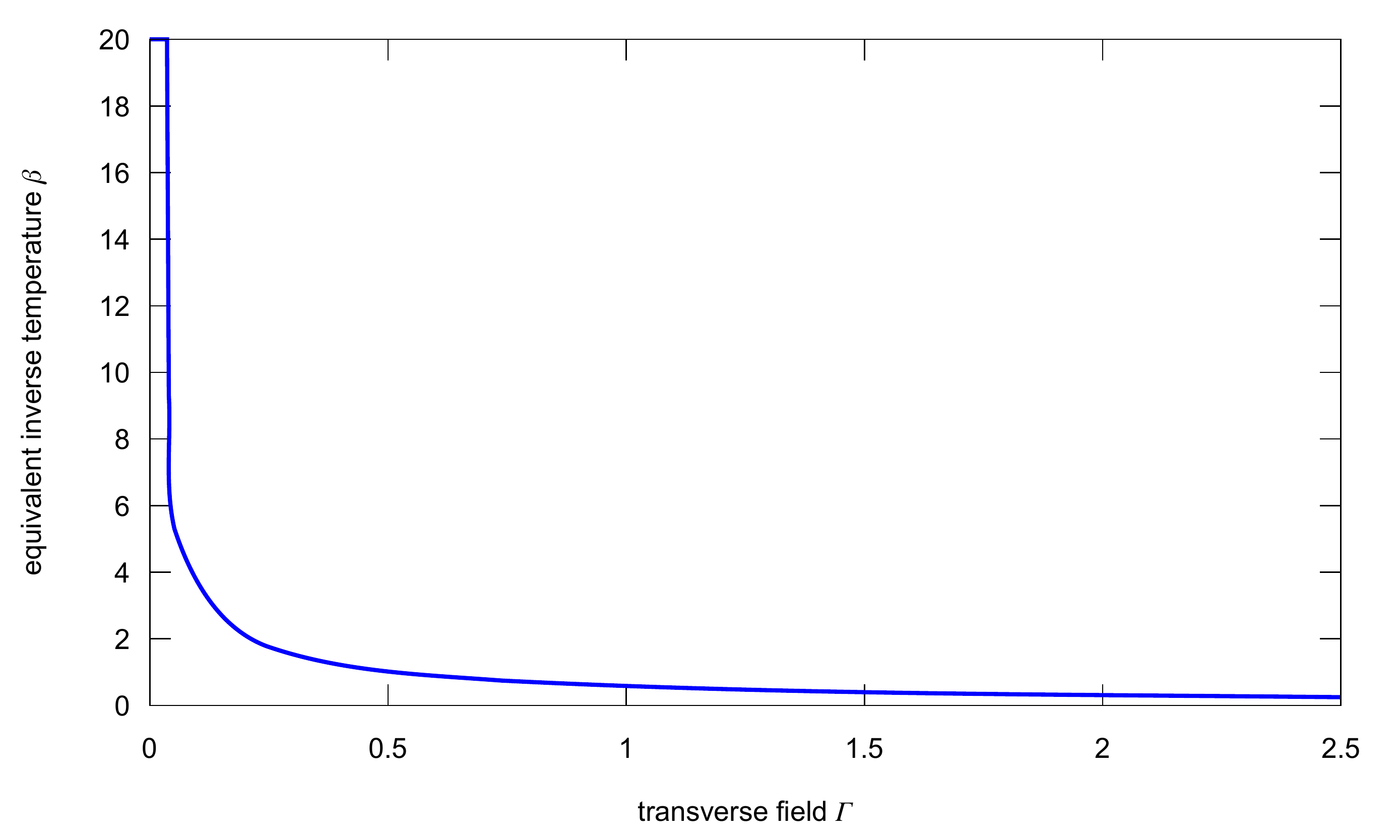}\caption{\label{fig:equiv_beta}The curve $\beta_{\mathrm{equiv}}\left(\Gamma\right)$
for $\alpha=0.4$ corresponding to a quantum system at $\beta=20$.}
\end{figure}

The results for SA presented in Fig.~\ref{fig:sim_vs_theory} used
an annealing protocol in $\beta$ designed to make a direct comparison
to QA: we found analytically a curve $\beta_{\mathrm{equiv}}\left(\Gamma\right)$
such that the classical equilibrium energy would be equal to the longitudinal
component of the quantum system energy, eq.~(\ref{eq:E-QA}). The
classical equilibrium energy was computed from the equations in ref.~\cite{krauth-mezard}.
The result is shown in Fig.~\ref{fig:equiv_beta}. The vertical jump
to $\beta=20$ is due to the transition mentioned in sec.~\ref{subsec:small-Gamma-n0};
as shown in Fig.~\ref{fig:sim_vs_theory}, the SA protocol in the
regime we tested gets stuck well before this transition.

The SA annealing protocol thus consisted in setting $\beta=\beta_{\mathrm{equiv}}\left(\Gamma\right)$
and decreasing linearly $\Gamma$ from $2.5$ to $0$, like for the
QA case. We fixed the total number of spin flip attempts at $\tau N\cdot10^{4}$
and used $\tau=4,8,16$; as for the QA case, the annealing process
was divided in $30\tau$ steps.

Other more standard annealing protocols (e.g. linear or exponential
or logarithmic) yielded very similar qualitative results, as expected
from the analysis of ref.~\cite{horner1992dynamics}.

\section{Additional numerical results on the annealing processes}

\subsection{Additional comparisons between theory and simulations}

\begin{figure}
\centering{}\includegraphics[width=0.5\columnwidth]{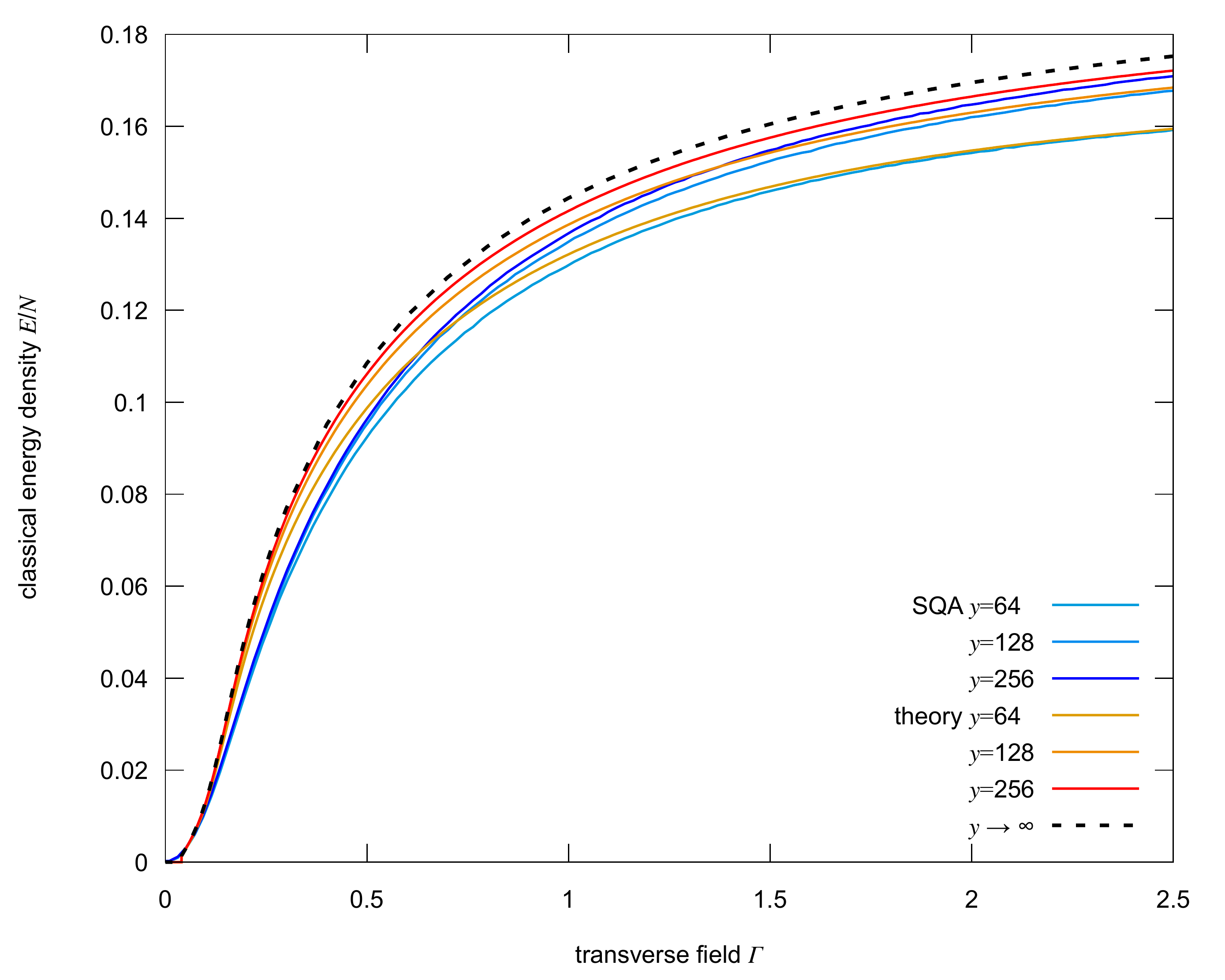}\caption{\label{fig:sim_vs_theory_y}Comparison between theory and simulations
for the average classical energy density at different values of $y$.
The three simulations curves (depicted in shades of blue) are the
same shown in Fig.~\ref{fig:sim_vs_theory}. At large $\Gamma$,
each of them is in good agreement with its corresponding analytical
curve (depicted in shades of red/orange). All the curves basically
coalesce at small $\Gamma$. In the intermediate regime, the theory
and the simulations exhibit a discrepancy, due to the static approximation
used in computing the analytical curves.}
\end{figure}

Fig.~\ref{fig:sim_vs_theory} compares the result of Monte Carlo
simulations with the theoretical predictions for the classical component
of the energy, eq.~(\ref{eq:E-QA}), and the transverse overlap,
eq\@.~(\ref{eq:q1}). Fig.~\ref{fig:sim_vs_theory_y} compares
the same simulation results with the analytical curves at finite $y$
instead. This shows a relatively small systematic offset (due to the
static approximation) at intermediate values of $\Gamma$, while the
agreement is good at both large and small $\Gamma$.

\begin{figure}
\centering{}\includegraphics[width=0.5\columnwidth]{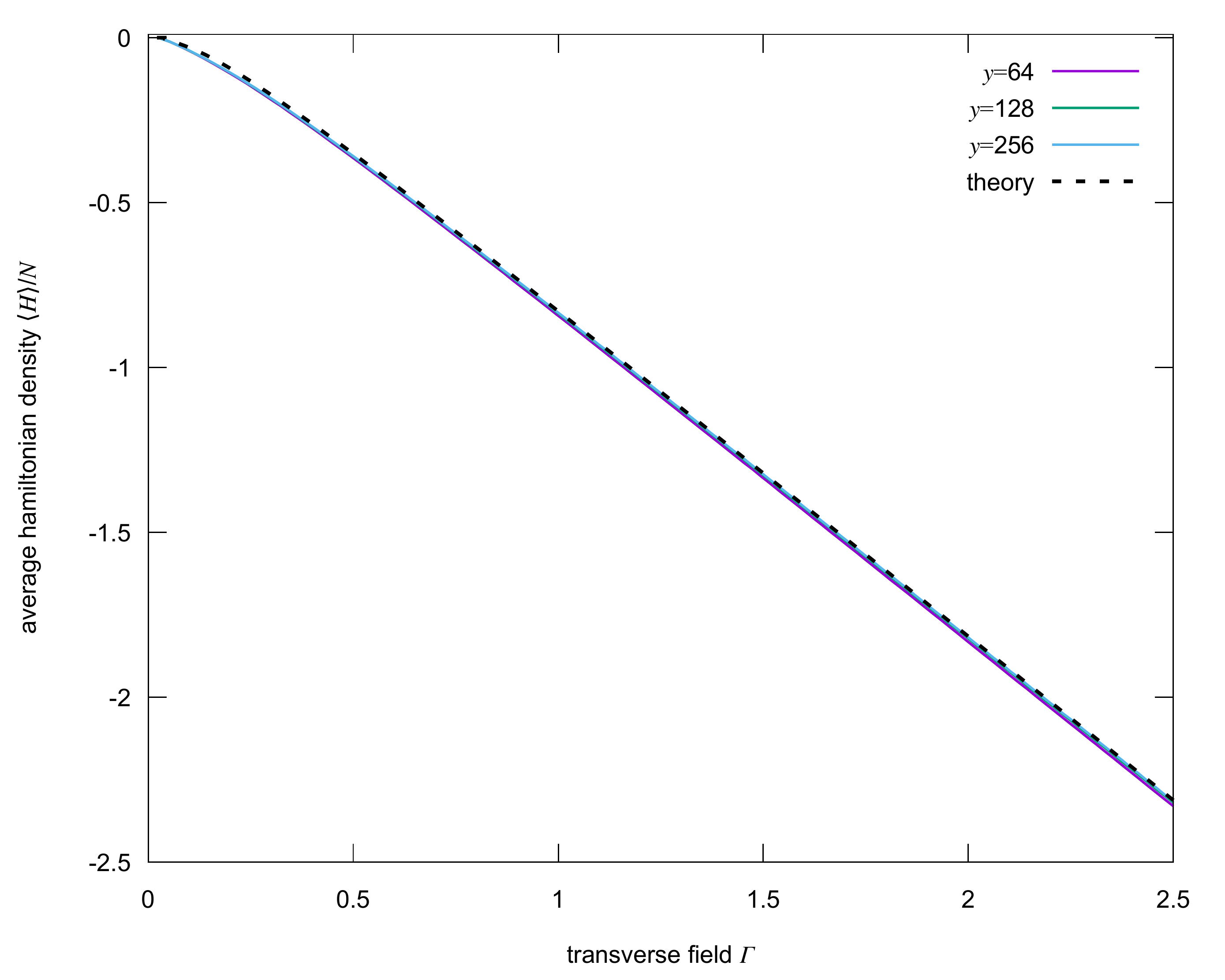}\caption{\label{fig:sim_vs_theory_H}Comparison between theory and simulations
for the average of the Hamiltonian density, eq.~(\ref{eq:H-QA})
divided by $N$. Same data as Fig.~\ref{fig:sim_vs_theory}. The
numerical curves are very close to the theoretical one at this level
of detail. A close inspection reveals that the agreement improves
with increasing $y$.}
\end{figure}

Fig.~\ref{fig:sim_vs_theory_H} shows the comparison with the $y\to\infty$
curve for the expectation of the full quantum Hamiltonian, eq.~(\ref{eq:H-QA}),
using the same data. The agreement is remarkable, and a close inspection
reveals that the curves from the simulation tend towards the theoretical
one as $y$ increases, i.e.~in the quantum limit.

\subsection{Experiments with two-layer networks\label{subsec:committee}}

We performed additional experiments using two-layer fully-connected
binary networks, the so-called committee machines. Previous results
obtained with the robust-ensemble measure \cite{baldassi_unreasonable_2016}
showed that this case is quite similar to that of single layer networks.
In particular, standard Simulated Annealing suffers from an exponential
slow-down as the system size increases even moderately, while algorithms
that are able to target the dense states do not suffer from the trapping
in meta-stable states. Indeed, we found the latter feature to be true
in the quantum annealing scenario.

The model in this case is defined by a modified energy function (cf.~eq.~(\ref{eq:E-1})):
\begin{equation}
E\left(\sigma\right)=\sum_{\mu=1}^{\alpha N}\Theta\left(-\sum_{k=1}^{K}\textrm{sgn}\sum_{j=1}^{N/K}\xi_{j}^{\mu}\sigma_{kj}\right)\label{eq:E-comm}
\end{equation}
where now the $N$ spin variables are divided in groups of $K$ hidden
units, and consequently the spin variables $\sigma_{kj}$ have two
indices, $k=1,\dots,K$ for the hidden unit and $j=1,\dots,N/K$ for
the input. Notice that the input size is reduced $K$-fold with respect
to the previous case. The output of these machines is simply decided
by the majority of the outputs of the individual units, and the energy
still counts the number of errors. The Suzuki-Trotter transformation
proceeds in exactly the same way as for the previous cases.

\begin{figure}
\centering{}\includegraphics[width=0.5\columnwidth]{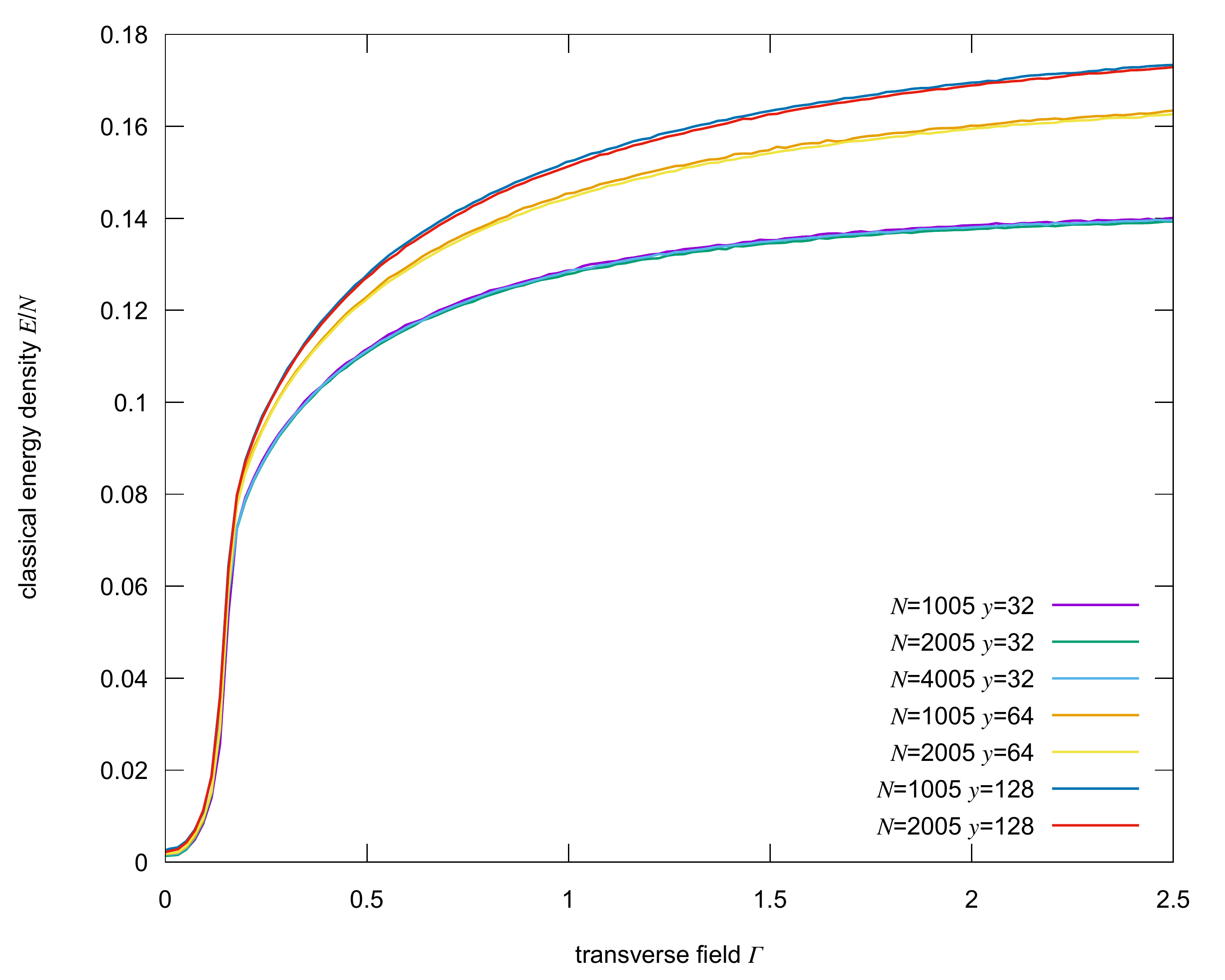}\caption{\label{fig:committee}Energy density eq.~(\ref{eq:E-comm}) as a
function of the transverse field $\Gamma$ for the two-layer binary
committee machine model with $K=5$ at $\alpha=0.4$ and $\beta=20$,
with different values of $N$ and $y$, and using $\tau=4$ in the
overall running time (number of spin flip attempts) set as $yN\tau\cdot10^{4}$.
Each curve is averaged over $15$ samples.}
\end{figure}

Like for the single-layer case, we tested the case of $\alpha=0.4$
at $\beta=20$, and we used $K=5$ units. We tested different values
of $N=1005,2005,4005$ with different values of the Trotter replicas
$y=32,64,128$ (only $y=32$ for $N=4005$) at a fixed overall running
time of $yN\tau\cdot10^{4}$ spin flip attempts, with $\tau=4$ (cf.~Fig.~\ref{fig:sim_vs_theory}).
The MC algorithm and the annealing protocols were also unchanged.
The results are shown in Fig.~\ref{fig:committee}: all these tests
produce curves which are almost indistinguishable at this level of
detail for different $N$, and that seemingly tend to converge to
some limit curve for increasing $y$ (while being almost overlapping
at small transverse field $\Gamma$), consistently with the single-layer
scenario.

\section{Real-time Quantum Annealing on small samples\label{sec:Real-time}}

\subsection{Numerical methods}

Computing the evolution of the system under Quantum Annealing amounts
at solving the time-dependent Schrödinger equation for the system
\begin{equation}
\frac{\partial}{\partial t}\left|\psi\left(t\right)\right\rangle =-i\hat{H}\left(t\right)\left|\psi\left(t\right)\right\rangle 
\end{equation}
where we set $\hslash=1$ for simplicity. In our case, the time dependence
of the Hamiltonian $H$ comes in through the varying transverse magnetic
field $\Gamma\left(t\right)$. We assume that $\Gamma$ varies linearly
with time between some starting value $\Gamma_{0}$ and $0$, in a
total time $t_{\max}$. Therefore, the final Hamiltonian is reduced
to the purely classical case, $\hat{H}\left(t_{\max}\right)=E$.

In the following, we will always work in the basis of the final Hamiltonian,
in which every eigenvector $\left|\sigma\right\rangle $ corresponds
to a configuration $\sigma\in\left\{ -1,+1\right\} ^{N}$ of the spins
in the $z$ direction. Therefore, we represent $\left|\psi\left(t\right)\right\rangle $
with a complex-valued vector of length $2^{N}$ with entries $\left\langle \sigma|\psi\left(t\right)\right\rangle $;
similarly, the $\hat{H}\left(t\right)$ operator is represented by
a matrix of size $2^{N}\times2^{N}$, $H\left(\sigma,\sigma^{\prime}\right)=\left\langle \sigma\left|\hat{H}\right|\sigma^{\prime}\right\rangle $.
The structure of this matrix is very sparse: the diagonal elements
$H\left(\sigma,\sigma\right)$ correspond to the classical energies
$E\left(\sigma\right)$, while the only non-zero diagonal elements
are those elements $H\left(\sigma,\sigma^{\prime}\right)$ such that
$\sigma$ and $\sigma^{\prime}$ are related by a single spin flip,
in which case the value is $-\Gamma$.

In our simulations, the initial state $\left|\psi\left(0\right)\right\rangle $
was set to the ground state of the system at $\Gamma\to\infty$, i.e.~with
all the spins aligned in the $x$ direction; in our basis, this corresponds
to a uniform vector, $\left\langle \sigma|\psi\left(0\right)\right\rangle =N^{-\nicefrac{1}{2}}$
for all $\sigma$. We simulated the evolution of the system by the
short iterative Lanczos (SIL) method \cite{schneider2016chapter}:
we compute the evolution at fixed $\Gamma$ for a short time interval
$\Delta t$, then lower $\Gamma$ by a small fixed amount $\Delta\Gamma$,
and iterate. The total evolution time is thus $t_{\max}=\frac{\Gamma_{0}}{\Delta\Gamma}\Delta t$.
Numerical accuracy can be verified by scaling both these steps by
a fixed amount and observing no significant difference in the outcome.
The evolution is computed by the Lanczos algorithm with enough iterative
steps to ensure sufficient accuracy, as determined by observing that
increasing the number of steps does not change the outcomes significantly.
In the simulations presented here, we set $\Gamma_{0}=5$, $\Delta\Gamma=10^{-3}$
and $\Delta t=0.2$, and we used $10$ steps in the Lanczos iterations.

At the end of the annealing process, we could retrieve the final probability
distribution for each configuration of the spins as $p\left(\sigma\right)=\left|\left\langle \sigma|\psi\left(t_{\max}\right)\right\rangle \right|^{2}$.

\subsection{Sample selection}

Given the exponential scaling with $N$ of the SIL algorithm, simulations
are necessarily restricted to small values of $N$. We used $N=21$.
At these system sizes, there is a very large sample-to-sample variability.
Furthermore, the energy barriers are generally small enough for the
classical Simulated Annealing to perform well.

In order to obtain small but challenging samples, in which we could
also study the structure of the solutions, we proceeded as follows:
we extracted at random $450$ samples with $P=17$ patterns each (corresponding
to $\alpha\simeq0.81$, close to the critical value of $0.83$ which
is valid for large systems), and selected those which had at least
a certain minimum number of solutions (note that in such small systems
we can easily enumerate all of the $2^{21}\simeq2\cdot10^{6}$ configurations
and check their energy). We arbitrarily chose $21$ solutions as the
threshold. We then ran both Simulated Annealing with a fast schedule
(with $\tau=1)$ and Simulated Quantum Annealing with $\tau=1$ and
a large number of Trotter replicas ($y=512$), and selected those
samples in which SA failed while SQA succeeded. This left us with
$20$ samples, which we then analyzed in detail and over which we
performed the real-time QA simulations.

\subsubsection{Randomized samples}

For each of the selected samples, we generated a corresponding randomized
version by permuting randomly the values of the energy associated
to each configuration. This procedure maintains unaltered the spectrum
of the energies (so that for example the classical Boltzmann distribution
at thermodynamic equilibrium remains unchanged), but completely destroys
the geometric features of the energy landscape. We used these randomized
samples as a benchmark against the measurements performed in our analysis.

\subsection{Analysis}

\subsubsection{QA vs SQA}

\begin{figure}
\centering{}\includegraphics[width=0.9\textwidth]{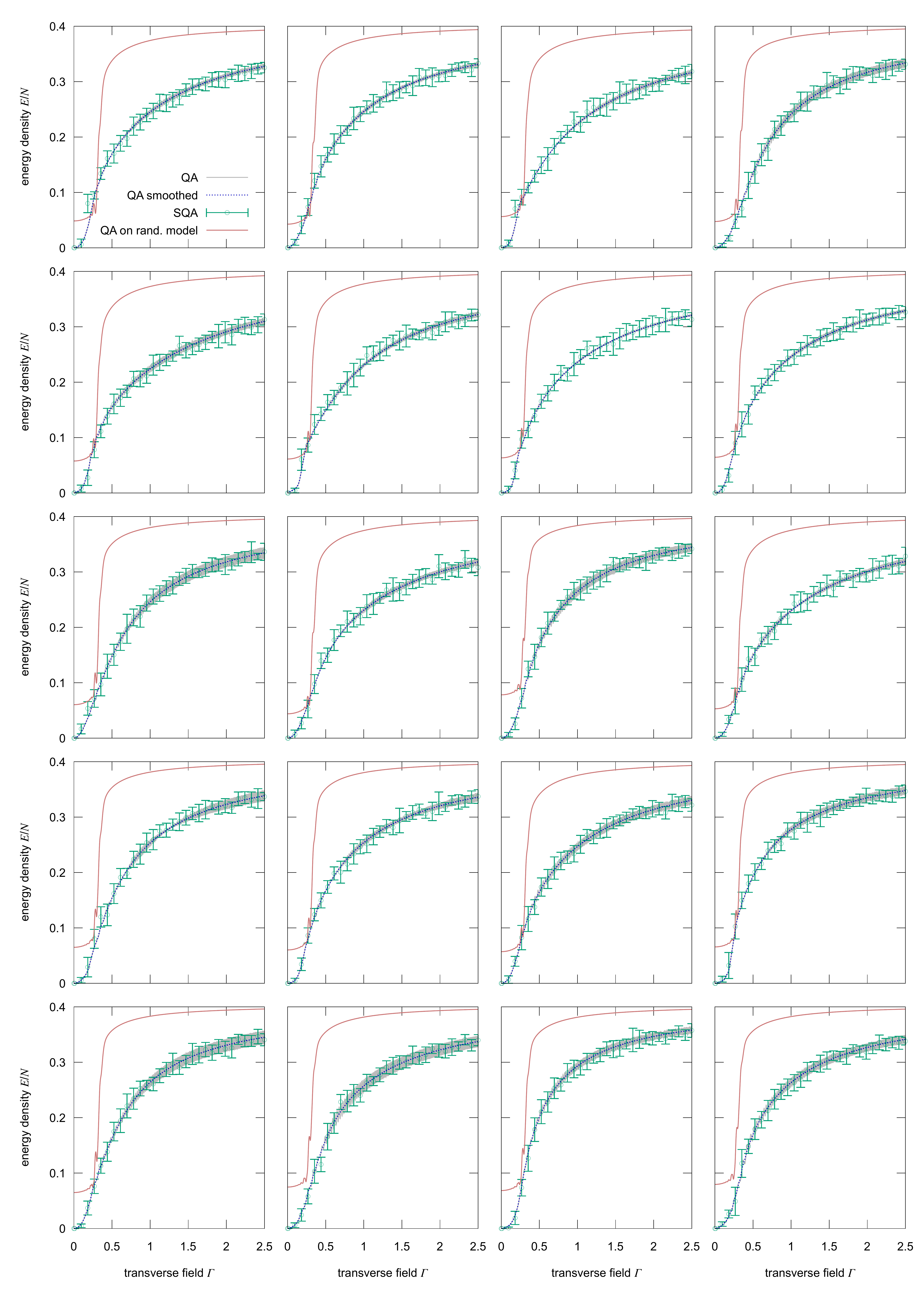}\caption{\label{fig:QA_vs_SQA-multi}Comparison between real-time QA and Monte
Carlo SQA for $20$ small samples with $N=21$. The figures show the
mean value of the longitudinal energy density $\left\langle E\right\rangle /N$
as a function of the transverse field $\Gamma$. The agreement between
the two algorithms is quite remarkable. The QA curves (shown in gray)
display some oscillatory behavior (not visible at this level of zoom
because the oscillations are fast) which however always tends to die
out as $\Gamma$ goes to $0$. The dotted blue curves show a smoothing
of these oscillations. The red curves show the results of the same
annealing process on a randomized version of the corresponding sample
(see text for details).}
\end{figure}

We compared the results of real-time QA with the SQA Monte Carlo results,
analyzing each of the $20$ selected samples individually. In particular,
we compared the values of the average longitudinal energy as a function
of $\Gamma$ for the two algorithms. As shown in Fig.~\ref{fig:QA_vs_SQA-multi},
the agreement is excellent, and the system always gets very close
to zero energy. In the same figure, we show that the same annealing
protocol however gives substantially different (and rather worse)
results on the randomized samples, reflecting the fact that the geometrical
features of the landscape are crucial (we verified on a few cases
that the results on the randomized samples could be improved by slowing
down the annealing process, but we could not get to the same results
as for the original systems even with a $100$-fold increase in total
time). Note that the sample-to-sample variability in these curves
appears to be fairly small due to our sample filtering process; we
verified in a preliminary analysis that the agreement is generally
excellent also without the filtering conditions, e.g.~on instances
that have no solutions at all.

\subsubsection{Other measurements}

We also performed a number of measurements on the final configuration
reached by the QA algorithm (both for the original samples and the
randomized ones) and studied the properties of the final probability
distribution $p\left(\sigma\right)$. These are the quantities that
we computed, reported in table~\ref{tab:QA_results}:
\begin{itemize}
\item The average value of the energy $\left\langle E\right\rangle =\sum_{\sigma}E\left(\sigma\right)p\left(\sigma\right)$.
\item The probability of finding a solution $P_{\mathrm{SOL}}=\sum_{\sigma:E\left(\sigma\right)=0}p\left(\sigma\right)$.
\item The probability and the energy of the most probable configuration,
$p\left(\sigma^{\star}\right)$ and $E\left(\sigma^{\star}\right)$,
where $\sigma^{\star}=\arg\max_{\sigma}p\left(\sigma^{\star}\right)$.
\item The inverse participation ratio $\mathrm{IPR}=\sum_{\sigma}p\left(\sigma\right)^{2}$,
to assess the concentration of the final distribution. (Qualitatively
analogous results are obtained using the Shannon entropy.) This measure
however does not take into account the geometric structure of the
distribution: for instance, if $p\left(\sigma\right)$ were non-zero
on just to configurations $\sigma^{1}$ and $\sigma^{2}$, the $\mathrm{IPR}$
would be very high, but it would not be able to discriminate between
the cases in which $\sigma^{1}$and $\sigma^{2}$ are close to each
other or far apart in Hamming distance.
\item The mean distance between configurations, defined as $\bar{d}=\sum_{\sigma,\sigma^{\prime}}p\left(\sigma\right)p\left(\sigma^{\prime}\right)d\left(\sigma,\sigma^{\prime}\right)$,
where $d\left(\sigma,\sigma^{\prime}\right)$ is the normalized Hamming
distance between configurations. This measure is useful since it reflects
the geometric features of the final measure: it can only be low if
the mass of the probability is concentrated spatially (in particular,
it is zero if and only if $p\left(\sigma\right)$ is a delta function).
\end{itemize}
As can be seen from the table, the results are generally in agreement
with the qualitative picture described in the main text, especially
when compared to the randomized benchmark: the system is able to reach
very low energies $\left\langle E\right\rangle $, the probability
of solving the problem $P_{\mathrm{SOL}}$ is very high, the measure
is rather concentrated on a few good configurations and those configurations
are close to each other (high $\mathrm{IPR}$, low $\bar{d}$).

For the original samples only, we also looked at the final configuration
from the Monte Carlo SQA process, $\sigma_{\mathrm{SQA}}$, and computed
its ranking according to $p\left(\sigma\right)$, which we denoted
as $r_{\mathrm{SQA}}$. A ranking of $1$ implies $\sigma_{\mathrm{SQA}}=\sigma^{\star}$.
All rankings are very small, the largest ones generally corresponding
to samples with the largest number of solutions and the less concentrated
distributions. This further attests to the good agreement between
the QA and SQA processes.

\begin{table}
\begin{raggedright}
{\footnotesize{}}%
\begin{tabular}{|c|c|c|c|c|c|c|c|c|}
\hline 
\multirow{2}{*}{{\footnotesize{}sample}} & \multirow{2}{*}{{\footnotesize{}\#sol}} & \multicolumn{6}{c}{\textbf{\footnotesize{}Original samples}} & \tabularnewline
\cline{3-9} 
 &  & {\footnotesize{}$P_{\mathrm{SOL}}$} & {\footnotesize{}$\left\langle E\right\rangle /N$} & {\footnotesize{}$E\left(\sigma^{\star}\right)$} & {\footnotesize{}$p\left(\sigma^{\star}\right)$} & {\footnotesize{}$\mathrm{IPR}$} & {\footnotesize{}$\bar{d}$} & {\footnotesize{}$r_{\mathrm{SQA}}$}\tabularnewline
\hline 
\hline 
{\footnotesize{}1} & {\footnotesize{}80} & {\footnotesize{}0.997873} & {\footnotesize{}0.007283} & {\footnotesize{}0} & {\footnotesize{}0.091049} & {\footnotesize{}0.052860} & {\footnotesize{}0.161271} & {\footnotesize{}2 }\tabularnewline
\hline 
{\footnotesize{}2} & {\footnotesize{}152} & {\footnotesize{}0.999345} & {\footnotesize{}0.005566} & {\footnotesize{}0} & {\footnotesize{}0.077510} & {\footnotesize{}0.035271} & {\footnotesize{}0.171339} & {\footnotesize{}30}\tabularnewline
\hline 
{\footnotesize{}3} & {\footnotesize{}25} & {\footnotesize{}0.998965} & {\footnotesize{}0.009524} & {\footnotesize{}0} & {\footnotesize{}0.260645} & {\footnotesize{}0.132882} & {\footnotesize{}0.116819} & {\footnotesize{}5 }\tabularnewline
\hline 
{\footnotesize{}4} & {\footnotesize{}129} & {\footnotesize{}0.999449} & {\footnotesize{}0.005236} & {\footnotesize{}0} & {\footnotesize{}0.163546} & {\footnotesize{}0.067067} & {\footnotesize{}0.121317} & {\footnotesize{}18}\tabularnewline
\hline 
{\footnotesize{}5} & {\footnotesize{}41} & {\footnotesize{}0.989324} & {\footnotesize{}0.022258} & {\footnotesize{}0} & {\footnotesize{}0.295728} & {\footnotesize{}0.149694} & {\footnotesize{}0.109064} & {\footnotesize{}1 }\tabularnewline
\hline 
{\footnotesize{}6} & {\footnotesize{}24} & {\footnotesize{}0.964274} & {\footnotesize{}0.041914} & {\footnotesize{}0} & {\footnotesize{}0.287578} & {\footnotesize{}0.169719} & {\footnotesize{}0.120534} & {\footnotesize{}1 }\tabularnewline
\hline 
{\footnotesize{}7} & {\footnotesize{}24} & {\footnotesize{}0.998103} & {\footnotesize{}0.010600} & {\footnotesize{}0} & {\footnotesize{}0.304858} & {\footnotesize{}0.150627} & {\footnotesize{}0.150411} & {\footnotesize{}3 }\tabularnewline
\hline 
{\footnotesize{}8} & {\footnotesize{}40} & {\footnotesize{}0.999237} & {\footnotesize{}0.006526} & {\footnotesize{}0} & {\footnotesize{}0.165771} & {\footnotesize{}0.078460} & {\footnotesize{}0.168933} & {\footnotesize{}3 }\tabularnewline
\hline 
{\footnotesize{}9} & {\footnotesize{}48} & {\footnotesize{}0.999253} & {\footnotesize{}0.005652} & {\footnotesize{}0} & {\footnotesize{}0.146795} & {\footnotesize{}0.070516} & {\footnotesize{}0.122878} & {\footnotesize{}2 }\tabularnewline
\hline 
{\footnotesize{}10} & {\footnotesize{}149} & {\footnotesize{}0.999120} & {\footnotesize{}0.009842} & {\footnotesize{}0} & {\footnotesize{}0.166722} & {\footnotesize{}0.084118} & {\footnotesize{}0.129071} & {\footnotesize{}16}\tabularnewline
\hline 
{\footnotesize{}11} & {\footnotesize{}27} & {\footnotesize{}0.999517} & {\footnotesize{}0.003968} & {\footnotesize{}0} & {\footnotesize{}0.292468} & {\footnotesize{}0.161192} & {\footnotesize{}0.101294} & {\footnotesize{}1 }\tabularnewline
\hline 
{\footnotesize{}12} & {\footnotesize{}69} & {\footnotesize{}0.999140} & {\footnotesize{}0.009151} & {\footnotesize{}0} & {\footnotesize{}0.325233} & {\footnotesize{}0.151204} & {\footnotesize{}0.081271} & {\footnotesize{}12}\tabularnewline
\hline 
{\footnotesize{}13} & {\footnotesize{}47} & {\footnotesize{}0.999550} & {\footnotesize{}0.004662} & {\footnotesize{}0} & {\footnotesize{}0.424602} & {\footnotesize{}0.230421} & {\footnotesize{}0.062514} & {\footnotesize{}3 }\tabularnewline
\hline 
{\footnotesize{}14} & {\footnotesize{}56} & {\footnotesize{}0.999123} & {\footnotesize{}0.006101} & {\footnotesize{}0} & {\footnotesize{}0.316729} & {\footnotesize{}0.151897} & {\footnotesize{}0.136677} & {\footnotesize{}8 }\tabularnewline
\hline 
{\footnotesize{}15} & {\footnotesize{}54} & {\footnotesize{}0.999430} & {\footnotesize{}0.005455} & {\footnotesize{}0} & {\footnotesize{}0.192050} & {\footnotesize{}0.100316} & {\footnotesize{}0.132264} & {\footnotesize{}1 }\tabularnewline
\hline 
{\footnotesize{}16} & {\footnotesize{}28} & {\footnotesize{}0.994853} & {\footnotesize{}0.010547} & {\footnotesize{}0} & {\footnotesize{}0.175557} & {\footnotesize{}0.115669} & {\footnotesize{}0.184412} & {\footnotesize{}3 }\tabularnewline
\hline 
{\footnotesize{}17} & {\footnotesize{}49} & {\footnotesize{}0.999546} & {\footnotesize{}0.004504} & {\footnotesize{}0} & {\footnotesize{}0.344131} & {\footnotesize{}0.169363} & {\footnotesize{}0.122593} & {\footnotesize{}5 }\tabularnewline
\hline 
{\footnotesize{}18} & {\footnotesize{}28} & {\footnotesize{}0.999361} & {\footnotesize{}0.006120} & {\footnotesize{}0} & {\footnotesize{}0.359053} & {\footnotesize{}0.187311} & {\footnotesize{}0.103266} & {\footnotesize{}6 }\tabularnewline
\hline 
{\footnotesize{}19} & {\footnotesize{}41} & {\footnotesize{}0.998693} & {\footnotesize{}0.004844} & {\footnotesize{}0} & {\footnotesize{}0.244020} & {\footnotesize{}0.111271} & {\footnotesize{}0.136971} & {\footnotesize{}1 }\tabularnewline
\hline 
{\footnotesize{}20} & {\footnotesize{}22} & {\footnotesize{}0.997396} & {\footnotesize{}0.007165} & {\footnotesize{}0} & {\footnotesize{}0.161562} & {\footnotesize{}0.108546} & {\footnotesize{}0.098991} & {\footnotesize{}4 }\tabularnewline
\hline 
\textbf{\footnotesize{}mean} & \textbf{\footnotesize{}56.65} & \textbf{\footnotesize{}0.996578} & \textbf{\footnotesize{}0.009346} & \textbf{\footnotesize{}0.0} & \textbf{\footnotesize{}0.239781} & \textbf{\footnotesize{}0.123921} & \textbf{\footnotesize{}0.126595} & \textbf{\footnotesize{}-}\tabularnewline
\hline 
\end{tabular}
\par\end{raggedright}{\footnotesize \par}
\begin{raggedright}
{\footnotesize{}}%
\begin{tabular}{|c|c|c|c|c|c|c|c|}
\hline 
\multirow{2}{*}{{\footnotesize{}sample}} & \multirow{2}{*}{{\footnotesize{}\#sol}} & \multicolumn{6}{c|}{\textbf{\footnotesize{}Randomized samples}}\tabularnewline
\cline{3-8} 
 &  & {\footnotesize{}$P_{\mathrm{SOL}}$} & {\footnotesize{}$\left\langle E\right\rangle /N$} & {\footnotesize{}$E\left(\sigma^{\star}\right)$} & {\footnotesize{}$p\left(\sigma^{\star}\right)$} & {\footnotesize{}$\mathrm{IPR}$} & {\footnotesize{}$\bar{d}$}\tabularnewline
\hline 
\hline 
{\footnotesize{}1} & {\footnotesize{}80} & {\footnotesize{}0.129571} & {\footnotesize{}1.018745} & {\footnotesize{}0} & {\footnotesize{}0.003140} & {\footnotesize{}0.000743} & {\footnotesize{}0.499356}\tabularnewline
\hline 
{\footnotesize{}2} & {\footnotesize{}152} & {\footnotesize{}0.241472} & {\footnotesize{}0.896787} & {\footnotesize{}0} & {\footnotesize{}0.004774} & {\footnotesize{}0.000834} & {\footnotesize{}0.499289}\tabularnewline
\hline 
{\footnotesize{}3} & {\footnotesize{}25} & {\footnotesize{}0.046472} & {\footnotesize{}1.184001} & {\footnotesize{}0} & {\footnotesize{}0.003105} & {\footnotesize{}0.000730} & {\footnotesize{}0.499035}\tabularnewline
\hline 
{\footnotesize{}4} & {\footnotesize{}129} & {\footnotesize{}0.210265} & {\footnotesize{}0.997248} & {\footnotesize{}0} & {\footnotesize{}0.003348} & {\footnotesize{}0.000814} & {\footnotesize{}0.499349}\tabularnewline
\hline 
{\footnotesize{}5} & {\footnotesize{}41} & {\footnotesize{}0.067973} & {\footnotesize{}1.211840} & {\footnotesize{}1} & {\footnotesize{}0.003300} & {\footnotesize{}0.000719} & {\footnotesize{}0.499451}\tabularnewline
\hline 
{\footnotesize{}6} & {\footnotesize{}24} & {\footnotesize{}0.042273} & {\footnotesize{}1.287633} & {\footnotesize{}1} & {\footnotesize{}0.002915} & {\footnotesize{}0.000712} & {\footnotesize{}0.499311}\tabularnewline
\hline 
{\footnotesize{}7} & {\footnotesize{}24} & {\footnotesize{}0.041953} & {\footnotesize{}1.331847} & {\footnotesize{}0} & {\footnotesize{}0.002760} & {\footnotesize{}0.000692} & {\footnotesize{}0.499459}\tabularnewline
\hline 
{\footnotesize{}8} & {\footnotesize{}40} & {\footnotesize{}0.068795} & {\footnotesize{}1.354127} & {\footnotesize{}1} & {\footnotesize{}0.002988} & {\footnotesize{}0.000711} & {\footnotesize{}0.499386}\tabularnewline
\hline 
{\footnotesize{}9} & {\footnotesize{}48} & {\footnotesize{}0.081738} & {\footnotesize{}1.269498} & {\footnotesize{}1} & {\footnotesize{}0.003505} & {\footnotesize{}0.000730} & {\footnotesize{}0.499100}\tabularnewline
\hline 
{\footnotesize{}10} & {\footnotesize{}149} & {\footnotesize{}0.230869} & {\footnotesize{}0.924125} & {\footnotesize{}0} & {\footnotesize{}0.003964} & {\footnotesize{}0.000809} & {\footnotesize{}0.499486}\tabularnewline
\hline 
{\footnotesize{}11} & {\footnotesize{}27} & {\footnotesize{}0.048923} & {\footnotesize{}1.642497} & {\footnotesize{}0} & {\footnotesize{}0.003199} & {\footnotesize{}0.000660} & {\footnotesize{}0.499143}\tabularnewline
\hline 
{\footnotesize{}12} & {\footnotesize{}69} & {\footnotesize{}0.111401} & {\footnotesize{}1.113771} & {\footnotesize{}1} & {\footnotesize{}0.002839} & {\footnotesize{}0.000738} & {\footnotesize{}0.499410}\tabularnewline
\hline 
{\footnotesize{}13} & {\footnotesize{}47} & {\footnotesize{}0.082780} & {\footnotesize{}1.366465} & {\footnotesize{}1} & {\footnotesize{}0.002932} & {\footnotesize{}0.000728} & {\footnotesize{}0.499171}\tabularnewline
\hline 
{\footnotesize{}14} & {\footnotesize{}56} & {\footnotesize{}0.096088} & {\footnotesize{}1.266256} & {\footnotesize{}1} & {\footnotesize{}0.002616} & {\footnotesize{}0.000741} & {\footnotesize{}0.499113}\tabularnewline
\hline 
{\footnotesize{}15} & {\footnotesize{}54} & {\footnotesize{}0.090438} & {\footnotesize{}1.201138} & {\footnotesize{}1} & {\footnotesize{}0.002936} & {\footnotesize{}0.000718} & {\footnotesize{}0.499126}\tabularnewline
\hline 
{\footnotesize{}16} & {\footnotesize{}28} & {\footnotesize{}0.051051} & {\footnotesize{}1.381148} & {\footnotesize{}0} & {\footnotesize{}0.002877} & {\footnotesize{}0.000689} & {\footnotesize{}0.499747}\tabularnewline
\hline 
{\footnotesize{}17} & {\footnotesize{}49} & {\footnotesize{}0.083151} & {\footnotesize{}1.364369} & {\footnotesize{}1} & {\footnotesize{}0.003114} & {\footnotesize{}0.000716} & {\footnotesize{}0.499719}\tabularnewline
\hline 
{\footnotesize{}18} & {\footnotesize{}28} & {\footnotesize{}0.050067} & {\footnotesize{}1.572161} & {\footnotesize{}1} & {\footnotesize{}0.003182} & {\footnotesize{}0.000680} & {\footnotesize{}0.499386}\tabularnewline
\hline 
{\footnotesize{}19} & {\footnotesize{}41} & {\footnotesize{}0.071946} & {\footnotesize{}1.439492} & {\footnotesize{}0} & {\footnotesize{}0.002916} & {\footnotesize{}0.000683} & {\footnotesize{}0.499483}\tabularnewline
\hline 
{\footnotesize{}20} & {\footnotesize{}22} & {\footnotesize{}0.041329} & {\footnotesize{}1.672801} & {\footnotesize{}0} & {\footnotesize{}0.002541} & {\footnotesize{}0.000656} & {\footnotesize{}0.499266}\tabularnewline
\hline 
\textbf{\footnotesize{}mean} & \textbf{\footnotesize{}56.65} & \textbf{\footnotesize{}0.094428} & \textbf{\footnotesize{}1.274798} & \textbf{\footnotesize{}0.5} & \textbf{\footnotesize{}0.003148} & \textbf{\footnotesize{}0.000726} & \textbf{\footnotesize{}0.499339}\tabularnewline
\hline 
\end{tabular}
\par\end{raggedright}{\footnotesize \par}
\caption{\label{tab:QA_results}Results for the $20$ small samples (original
an randomized) at the end of the QA process. The samples are the same
as for Figg.~\ref{fig:QA_vs_SQA-multi} and~\ref{fig:loc_entropy-multi},
where they are arranged in row-major order. The second column shows
the number of solutions; the other columns are described in the text.}
\end{table}

\subsubsection{Local entropies}

In order to assess whether the denser ground states were favored in
the final configuration with respect to more isolated solutions we
compared the mean local entropy curves weighted according to $p\left(\sigma\right)$
with those averaged over all the solutions. More precisely, we define
$C\left(n\right)$ as the set of the $n$ configurations with highest
probability, and $n_{w}$ as the number of configurations required
to achieve a cumulative probability of $t$, i.e.~the lowest $n$
such that $\sum_{\sigma\in C\left(n\right)}p\left(\sigma\right)\ge w$.
We also define $K\left(\sigma,d\right)$ as the number of solutions
at normalized Hamming distance from $\sigma$ lower or equal to $d$.
Then the mean local entropy curve weighted with $p$ is then defined
as:
\begin{equation}
\phi_{w}\left(d\right)=\frac{1}{N}\frac{\sum_{\sigma\in C\left(n_{w}\right)}p\left(\sigma\right)\log K\left(\sigma,d\right)}{\sum_{\sigma\in C\left(n_{w}\right)}p\left(\sigma\right)}.
\end{equation}

Denoting by $\mathcal{S}=\left\{ \sigma|E\left(\sigma\right)=0\right\} $
the set of all the solutions, we also compute the flat average of
the local entropies over $\mathcal{S}$:
\[
\phi_{\mathrm{SOL}}\left(d\right)=\frac{1}{N\left|\mathcal{S}\right|}\sum_{\sigma\in\mathcal{S}}\log K\left(\sigma,d\right).
\]

\begin{figure}
\centering{}\includegraphics[width=0.9\textwidth]{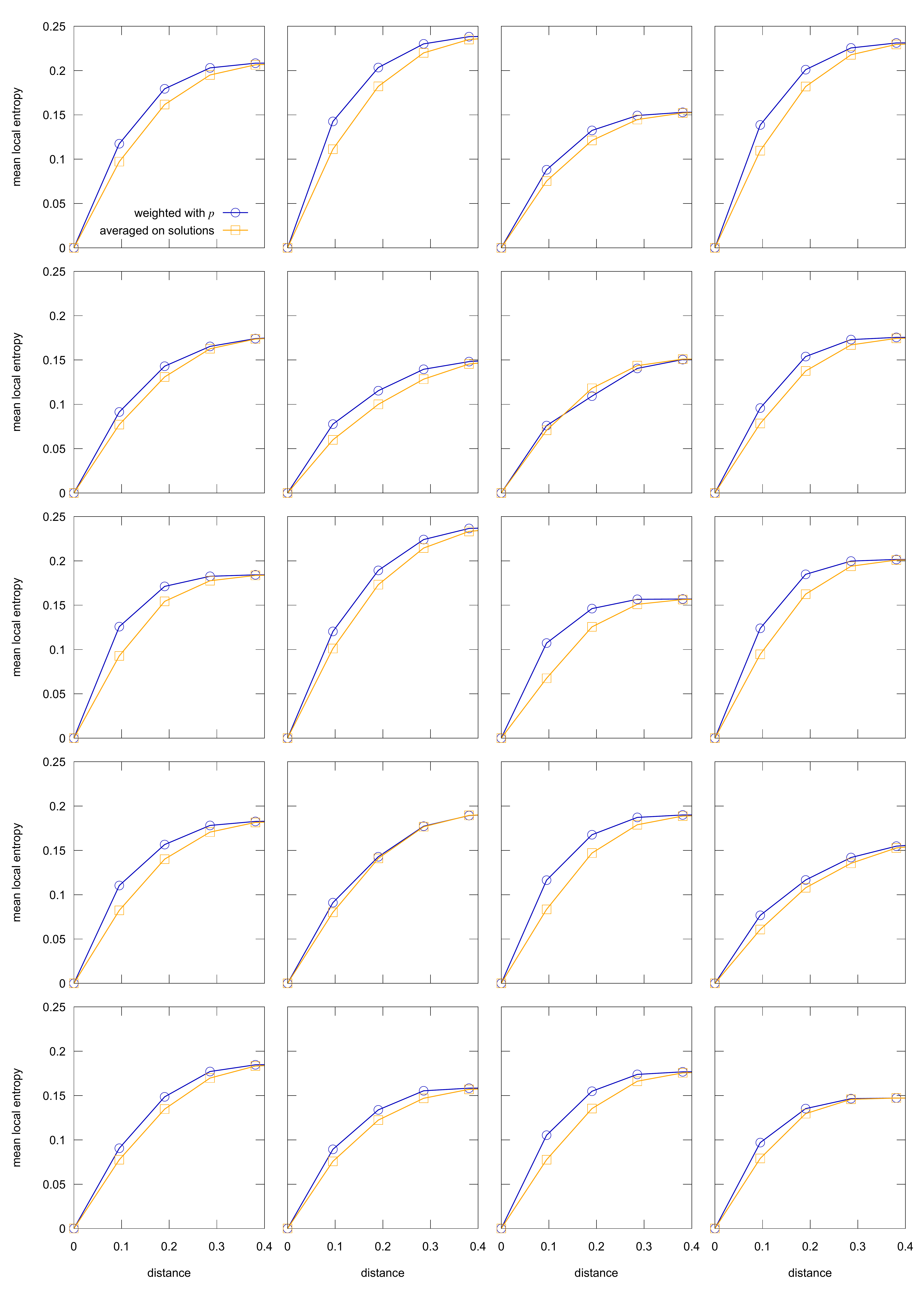}\caption{\label{fig:loc_entropy-multi}Comparison of the average local entropies
weighted according to the probability distribution obtained at the
end of the annealing process, with those obtained from a flat average
over all solutions, for $20$ different small samples with $N=21$.
The samples are the same as in Fig.~\ref{fig:QA_vs_SQA-multi}.}
\end{figure}

If $p$ concentrates on denser solutions, we expect that the $\phi_{w}$
curves should be generally higher than the $\phi_{\mathrm{SOL}}$
curves. Indeed, the results confirm this scenario, as shown in Fig.~\ref{fig:loc_entropy-multi},
where we used $w=0.9$. (This value ensured that $C\left(n_{w}\right)\subseteq\mathcal{S}$
for all samples and thus that all the local entropies are finite;
apart from this, the results are quite insensitive to the choice of
$w$.) Note that, in the limit of large system sizes, the $\phi_{\mathrm{SOL}}$
curves would be dominated by isolated solutions and display a gap
around zero distances; the fact that this is not visible in Fig.~\ref{fig:loc_entropy-multi}
is purely a finite size effect; the $\phi_{w}$ curves on the other
hand should be roughly comparable to those shown in Fig.~\ref{fig:en_landscapes}.

\subsubsection{Energy gaps}

\begin{figure}
\centering{}\includegraphics[width=0.9\textwidth]{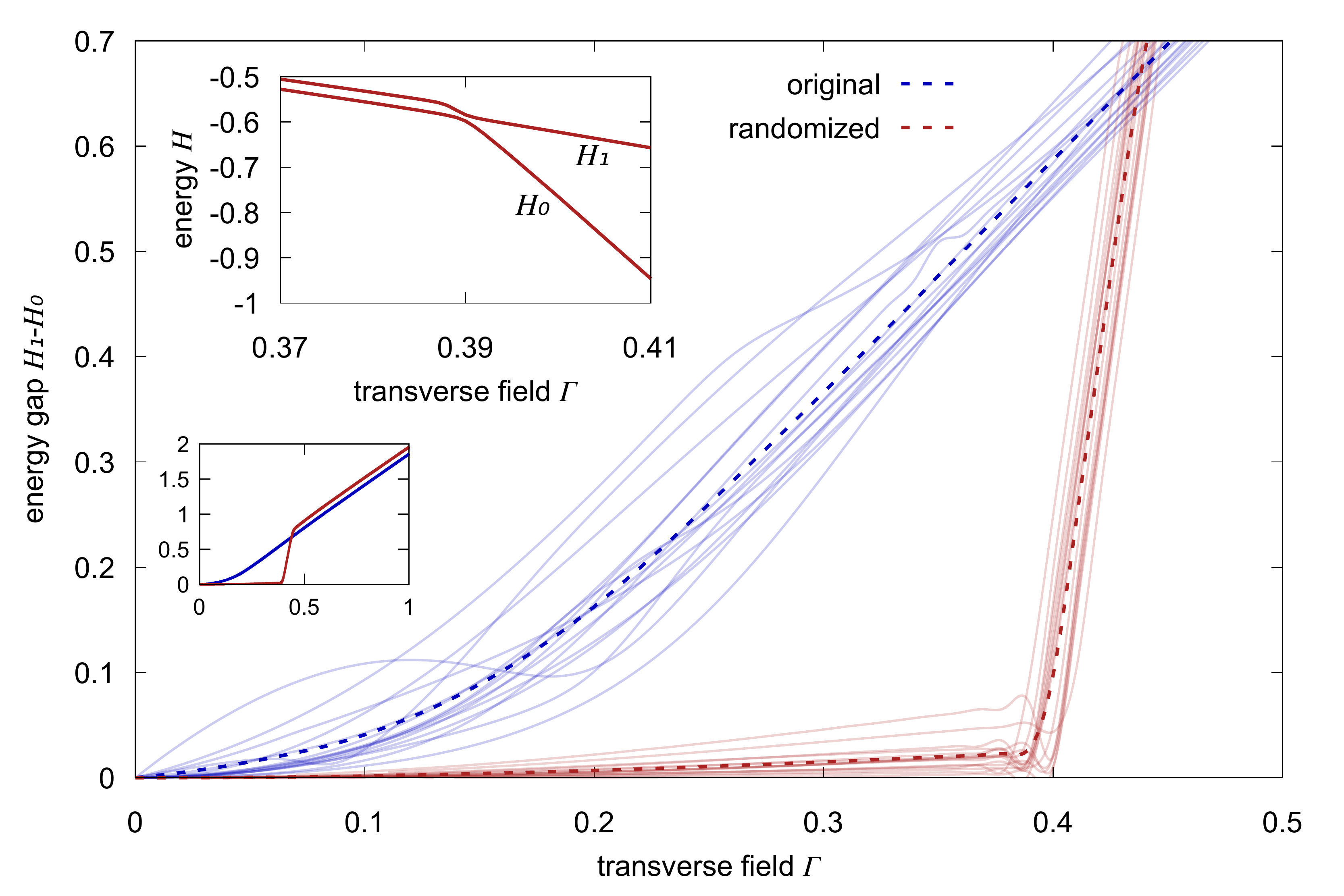}\caption{\label{fig:gaps}Energy gap between the ground state $H_{0}$ and
the first excited state $H_{1}$ as a function of the transverse field
$\Gamma$, for $20$ small samples with $N=21$ (same as in Fig.~\ref{fig:QA_vs_SQA-multi}).
The semi-transparent solid curves show the results for each individual
sample (blue: original; red: randomized), while the dashed lines are
averages. The behavior is qualitatively different for the two cases:
the randomized examples all display avoided crossings at $\Gamma\simeq0.4$
(the upper inset figure shows the two energy levels for one representative
example, enlarged around the relevant region); the original examples
on the other hand show no trace of avoided crossings (with one possible
exception) and are generally much higher. The lower inset shows the
same data as the main figure, but only the averages are plotted, and
the range is enlarged up to $\Gamma=1$.}
\end{figure}

As mentioned in the introduction of the main text, it is well known
that, according to the adiabatic theorem, the effectiveness of the
QA process depends on the relation between the rate of change of the
Hamiltonian and the size of the gap between the ground state of the
system $H_{0}$ and the first excited state $H_{1}$: smaller gaps
require a slower annealing process. Therefore, we performed a static
analysis of the energy spectrum of each of the $20$ samples at varying
$\Gamma$, and computed the gap $H_{1}-H_{0}$, comparing the results
with those for the randomized versions of the samples. The results
are shown in Fig.~\ref{fig:gaps}. For the original samples, the
gap only vanishes in the limit of $\Gamma\to0$ (which is expected
since the ground state at $\Gamma=0$ is degenerate). For the randomized
samples, on the other hand, the gap nearly closes at non-zero $\Gamma$,
displaying the characteristics of an ``avoided crossing'' (see the
figure upper inset), which is the type of phenomenon that is known
to hamper the performance of QA algorithms. Indeed, the values of
$\Gamma$ where these avoided crossings occur are precisely those
at which the mean value of $H$ found by the QA algorithm deviates
from the ground state $H_{0}$, thereby getting stuck as shown in
Fig.~\ref{fig:QA_vs_SQA-multi}.

\end{document}